\documentclass[prd,preprintnumbers,twocolumn,eqsecnum,floatfix,superscriptaddress,nofootinbib]{revtex4}

\usepackage[utf8]{inputenc}
\usepackage{graphicx}
\usepackage{dcolumn}
\usepackage{bm}
\usepackage[utf8]{inputenc}
\usepackage{bm}
\usepackage{latexsym,amssymb,amsmath,amsfonts}
\usepackage{graphicx}
\graphicspath{{plots/}}
\usepackage[usenames]{color}
\usepackage{array}
\usepackage[breaklinks, backref, colorlinks=true]{hyperref}

\def\MM{\textrm{MM}}

\newcommand{\vek}[1]{\boldsymbol{#1}}

\begin{document}
\preprint{IUCAA-05/2018, LIGO-P1800160}

\title{Hierarchical search strategy for the efficient detection of gravitational waves from non-precessing coalescing compact binaries with aligned-spins}

\author{Bhooshan~Uday~Varsha~Gadre}
\email{bug@iucaa.in}
\affiliation{Inter-University Centre for Astronomy and Astrophysics (IUCAA), Post Bag 4, Ganeshkhind, Pune 411007, India}

\author{Sanjit~Mitra}
\email{sanjit@iucaa.in}
\affiliation{Inter-University Centre for Astronomy and Astrophysics (IUCAA), Post Bag 4, Ganeshkhind, Pune 411007, India}

\author{Sanjeev~Dhurandhar}
\email{sanjeev@iucaa.in}
\affiliation{Inter-University Centre for Astronomy and Astrophysics (IUCAA), Post Bag 4, Ganeshkhind, Pune 411007, India}


\begin{abstract}

In the first two years of Gravitational Wave (GW) Astronomy, half a dozen compact binary coalescences (CBCs) have been detected. As the sensitivities and bandwidths of the detectors improve and new detectors join the network, many more sources are expected to be detected. The goal will not only be to find as many sources as possible in the data but to understand the dynamics of the sources much more precisely. Standard searches are currently restricted to a smaller parameter space which assumes aligned spins. Construction of a larger and denser parameter space, and optimising the resultant increase in false alarms, pose a serious computational challenge. We present here a two-stage hierarchical strategy to search for CBCs in  data from a network of detectors and demonstrate the computational advantage in real life scenario by introducing it in the standard {\tt PyCBC} pipeline with the usual restricted parameter space. With this strategy, in simulated data containing stationary Gaussian noise, we obtain a computational gain of $\sim 20$ over the flat search. In real data, we expect the computational gain up to a factor of few. This saving in the computational effort will, in turn, allow us to search for precessing binaries. Freeing up computation time for the regular analyses will provide more options to search for sources of different kinds and to fulfil the never-ending urge for extracting more science out of the data with limited resources.

\end{abstract}

\maketitle

\section{Introduction}
\label{intro}

During the first (O1) and the second (O2) observation runs, the twin LIGO (Laser Interferometric Gravitational-wave Observatory) detectors observed gravitational wave (GW) signals from 6 events with confidence - 5 mergers of binary black holes (BBHs) and one double neutron star coalescence. The neutron star coalescence had electromagnetic counterparts in almost every band and is even now being followed by many electromagnetic (radio) telescopes. For the last two of the observations of compact binary coalescences (CBCs), the data from the VIRGO detector also was used supplementing the LIGO~\cite{Abbott:2016blz,Abbott:2016nmj,Abbott:2017vtc, Abbott:2017gyy,Abbott:2017iws,Abbott:2017oio,GBM:2017lvd} data.  We soon expect to have a larger network of such interferometric detectors with KAGRA coming online soon, and LIGO-India following in few years~\cite{Iyer2011,Sathya2012,Akutsu:2017thy}. CBCs are perhaps going to be the most abundant sources for the current and next generation terrestrial interferometric GW detectors~\cite{Abbott:2016nhf}.

However, GW signals are usually buried deep into noisy interferometric strain data. To extract the signals from CBCs, where phase can be precisely modelled, the method of matched filtering is generally used~\cite{Sathyaprakash:1991mt,Dhurandhar:1992mw,Dhurandhar:1994mi} which is optimal in several ways.
The signal waveform for a particular set of parameters is obtained from the general theory of relativity by using various techniques involving analytical approximations, perturbation theory, numerical relativity, etc.\cite{Faye:2012we,Ajith,Taracchini:2012,Taracchini:2013rva,SpECwebsite,Husa:2015iqa}. The modelled signal is then cross-correlated with the inverse noise power weighted data from each of the detectors. This correlation is in fact the maximum likelihood estimator. If the signal with a loud enough correlation is simultaneously present in the pair of detectors with matching parameters, then we believe it as a real astrophysical signal, where the significance of detection needs to be estimated from the statistical properties of the data. This is a simplistic picture however of how coincident detection works. The difficulty lies in the fact that we do not know the signal parameters {\it a priori} and therefore a search must be carried out in the deemed parameter space.

For these searches, we assume quasi-circular orbits for the CBCs. For the circular orbits, the GW waveforms depend upon 15 parameters which can be split into two distinct classes: 8 intrinsic and 7 extrinsic. The intrinsic parameters are the component masses ($m_1$, $m_2$), individual spin angular momenta ($\vek s_1$, $\vek s_2$), sky location ($\theta$, $\phi$), luminosity distance ($d_L$), orbital inclination ($\mathcal{\iota}$), time and phase of coalescence ($t_c$, $\phi_c$) and polarisation angle ($\psi$). The dynamics of the source depends only upon the intrinsic parameters. We can model the generic GW signal in the source frame using essentially the intrinsic parameters and then transform it subsequently to the wave frame. For data analysis also, the intrinsic and extrinsic parameters are dealt with differently. One makes use of the symmetries in the signal model which makes the search computationally efficient. For example, we employ Fast Fourier Transform ({FFT}) to search over $t_c$ and use the waveforms with phases $0$ and $\pi/2$ as a basis to optimize the search over $\phi_c$ and then similarly deal with other extrinsic parameters in a quick way~\cite{Allen:2005fk,Pai:2000zt}. However, for the intrinsic parameters, we need to discretise the deemed parameter space. This set of GW signal waveforms at discrete points systematically sampled over the intrinsic  parameters is known as a template bank~\cite{Owen:1995tm,DalCanton:2017ala,Harry:2009ea,VanDenBroeck:2009gd,Roy:2017oul}. We then search for the signal by correlating all the templates in the bank with the detector data. To search for CBCs in current data from LIGO-Virgo detectors, typically few hundred thousand templates are needed to cover the parameter space densely enough for each data segment, demanding significant amount of computation.

The computational cost for a matched filtering search in the full parameter space is not feasible with the available resources. The current searches make a simplifying assumption to reduce the dimensionality of the parameter space---the spins of the binary components are assumed to be aligned with the orbital angular momentum. These non-precessing templates can detect a good part of the full parameter space when precessional effects are not dominant~\cite{Harry:2013tca}. With this set up, matched filter based LIGO pipelines use the template bank with minimal match (MM), the minimum value of scalar product between any two normalised templates,
of $0.97$. For a search up to a total mass of the 100 $M_{\odot}$, $\sim 250,000$ templates are required~\cite{DalCanton:2017ala}. As the low frequency sensitivity of the detectors improve, CBC signals in the detector band become longer with many more cycles. This demands increase in the template density in the parameter space. Further, better sensitivity at lower frequencies means that we can also observe the heavier binaries, resulting in extending the detectable parameter space. Both the effects together leads to at least few times increase in the non-precessing template bank.

Computation cost is orders of magnitude larger when searching for GW signals from precessing CBC systems. It has been shown that, even for the restricted parameter space of mass ratio less than 5, the 0.9 minimal match precessing template bank is more than 10 times larger than the corresponding non-precessing template bank with 0.97 minimal match~\cite{Harry:2016ijz}. Fortunately, precession of the binary becomes important only when masses are unequal and orbital inclination is not nearly face-on~\cite{Harry:2013tca}. Since less power in GW is emitted along the orbital plane, the chances of detection of such binaries have been low, which is why one could justify restricting the current matched filtering searches for CBC to dominant mode(s) of non-precessing signal models only~\cite{Capano:2016dsf,Babak:2012zx}. However, with progressively increasing sensitivities of the detectors and the addition of more detectors into the network, one can no longer afford to miss precessing binaries and the interesting science that they have to offer.
While there are claims that, through secular evolution, the component spins of the compact binaries are more likely to align or anti-align to the orbital angular momentum when they enter LIGO's sensitive band, sensitive searches for precessing binaries will again be necessary to test such claims via null detections. Such searches are clearly not feasible using the standard matched filtering scheme with available computing resources. This makes a strong case to develop cost reducing algorithms.

In general, due to the constant demand to extract more science out of a given set of data, computational costs could get very high and perhaps out of bounds given the current computational resources. The present matched filtering searches employ the coincident detection strategy, instead of the more detection efficient coherent strategy because the coherent strategy is significant computationally more expensive than the coincident searches~\cite{Macleod:2015jsa}. It is therefore very important to develop cost effective algorithms for matched-filter based searches, which will allow us to provide more computing resources to search for GWs from other astrophysical sources, e.g., from millisecond pulsars, and will enable us to perform more sophisticated searches, e.g., the precessing coherent search online which is the holy-grail of the CBC searches!

In this paper, we propose a hierarchical strategy to search for CBCs in data from a network of GW detectors to reduce the computational cost of the analysis. We demonstrate the benefit of this method using spin-aligned template banks. These banks have the advantage that there are fewer parameters over which the search needs to be carried out - there are only four intrinsic parameters to be reckoned with, the two masses and two component spins parallel to the orbital angular momentum. Also since the systems do not precess, the orbital inclination parameter $\mathcal{\iota}$ becomes redundant.

Current searches like GSTLAL, use singular value decomposition (SVD) like algorithms to numerically reduce the size of the non-precessing template banks which makes the matched filter part of the search computationally significantly cheaper but the reconstruction from the SVD basis to the actual binary template filtered output require extra computation~\cite{Cannon:2010qh}. Hence there is not significant saving in terms of computation as compared to direct search using the template bank. There are, still, some extra benefits of using {\tt PyCBC} like flat searches and speeding them up so that we can try to accommodate higher modes and precessing effects in the templates.
Here we introduce the hierarchical detection strategy to speed up the matched filtering search using the {\tt PyCBC} pipeline~\cite{Canton:2014ena,Usman:2015kfa,Nitz:2017svb,pycbc-software}. We only consider a 2-stage hierarchical search and compare it with the matched filtering search similar to what was used for the analysis of advanced LIGO's first observation run (O1) data. We find that we recover almost all the CBC signals with similar significance as the flat search but with almost {\it 20 times less computational cost} in simulated data containing Gaussian stationary noise. This result is obtained with some caveats which we  discuss in the text with more details.

The layout of the paper is as follows. In section~\ref{Pre}, we begin by describing earlier work that has been done on hierarchical strategies in the context of CBCs. Then we describe the flat search or the single stage search with matched filtering. In section~\ref{hierarchical search}, we briefly review the previous use of hierarchical algorithm and then we discuss the current implementation of the $2$-stage hierarchical search. Then in the section~\ref{hierarchical_result}, we compare the results of this implementation of the hierarchical search with the flat search using aLIGO like simulated data. Finally in section~\ref{discussion}, we summarise and discuss the future directions and also the procedure we would like to adopt in these strategies. The work that is presented in this paper has been with the aim of showing proof of concept. We are working on the optimization of the method which will make use of  real detector data.

\section{Preliminaries}
\label{Pre}


\subsection{The matched filter}
\label{mf}

The matched filter (MF) is noise weighted correlation of the modelled GW signal (the template) with the data. It is an optimal detection statistic (in the Neyman-Pearson sense), surrogate of the maximum likelihood statistic when the noise is stationary and Gaussian ~\cite{Creighton:2011zz,Sathyaprakash:1991mt,Dhurandhar:1994mi}. The mathematical form of the MF statistic, which is same as the SNR (usually denoted by $\rho$) for normalised templates, is maximised over phase of coalescence analytically and also all other parameters of the signal for non-precessing waveforms, and is given by,
\begin{equation}
\begin{split}
 \rho &\equiv \max_{\bm{\lambda}} \left( x, h \left(\bm{\lambda} \right) \right) \\
 ~ &=~ \max_{\bm{\lambda}}~\left( 4~ \text{Abs} ~\int_0^\infty \frac{\tilde{x}^{*}\left( f \right) \left( \tilde{h} + i \tilde{h} \right) \left( f; {\bm{\lambda}} \right)}{S_{n} \left( f \right)} ~\mathrm{d}f \right) \, ,
\end{split}
\label{eq:mf}
\end{equation}
where $x$ is the time series strain data, $h(\bm{\lambda})$ is the normalized expected GW signal for the set of binary source parameters given by $\bm{\lambda}$ and $S_n(f)$ is the noise power spectral density (PSD). The round brackets denote a scalar product on the space of data trains, which has been defined in Eq. (\ref{eq:mf}) in the Fourier domain. Also tilde ($\tilde{\ }$) above a quantity, denotes the Fourier space representation of the time series representation of the function. Because of the maximisation over phase in the MF, in stationary Gaussian noise, the detection statistic follows a Rayleigh probability distribution in absence of the signal and a Rician distribution when a signal is present in the data ~\cite{Mukhopadhyay:2009qh}. In general, we have no knowledge of the signal parameters $\bm{\lambda}$ and therefore we must search over the full parameter space to carry out the maximisation. The search over the time of coalescence is performed in a quick way by using Fast Fourier Transform (FFT). For the rest of parameters, namely, the intrinsic parameters, as discussed in the introduction, we require a template bank. The template bank is constructed with MM  of $0.97$. In the next subsection we now explain the concept of template bank.

\subsection{The template bank}
\label{bank}

The discrete sampling of the intrinsic parameters has to be done with due care. Otherwise we may miss out signals due to the loss of SNR because of the mismatch in the template and signal parameters. There can be many reasons for loss in SNR, mainly it is the phase mismatch which matters the most, which may be due to inaccurate modelling of the signal, etc. But one of the reasons is the mismatch due to the discrete nature of the template bank. As the templates are normalized, {\unexpanded{$\left( h \left(\bm{\lambda} \right), h \left(\bm{\lambda} \right) \right) = 1$}, a match between any of the two waveforms with slightly different parameters can be written as follows:
\begin{equation}
\begin{split}
 \mathcal{H} \left(\bm{\lambda},~ \bm{\lambda+\triangle \lambda} \right) &\equiv \left( h \left(\bm{\lambda} \right), h \left(\bm{\lambda+\triangle \lambda} \right) \right) \\
 ~&=~ 1 ~-~ ds^2 ~=~ 1 ~-~ g_{ab}\left(\bm{\lambda} \right) \triangle \lambda^a \triangle \lambda^b \, ,
\end{split}
 \end{equation}
where we have kept lowest order terms in $\bm{\triangle \lambda}$ and defined the metric $g_{ab}$ as:
\begin{equation}
 g_{ab}\left(\bm{\lambda} \right) ~=~ -\frac{1}{2} \left( h \left(\bm{\lambda} \right),~ \frac{\partial^2 h}{\partial \lambda^a \partial \lambda^b}\left(\bm{\lambda} \right) \right) \, .
\end{equation}
The distance $ds$ and template space metric $g_{ab}$ can be used to systematically place templates in the bank with a given value of \MM, provided $1 - \MM$ is small. Usually the mismatch $1 - \MM$ is chosen as $3 \%$~\cite{DalCanton:2017ala}, that is, $\MM = 0.97$, which corresponds to a maximum loss of 10\% of the astrophysical events within the detectable range. The metric can be analytically calculated for inspiral waveforms given by the post-Newtonian expansion. But here, we use the full IMR waveforms with non-precessing component spins in the search. For such waveforms, there is no sufficiently accurate analytic or semi-analytic  form of the metric which can be used to construct a geometric template bank. Therefore, the current searches use a different approach which employs stochastic methods in order to obtain the template bank ~\cite{Harry:2009ea,Roy:2017oul}, where the match is directly computed to obtain a stochastic placement of templates. If the the match is close to unity, then the metric is being used implicitly. If the match is not close to unity as in the case of the coarse bank as explained in Section~\ref{Formalism}, then the metric approximation fails.

A template bank depends on the PSD of noise present in the detector. However, when we have more than one detector, in general, we have to deal with more than one PSDs. However, it is convenient to have a common template bank, which facilitates the coincident detection approach~\cite{Capano:2016dsf,DalCanton:2017ala}. For the two LIGO detectors, we combine the two PSDs as a harmonic mean to construct a common effective template bank for the search.  As the strain noise from the LIGO detectors is neither stationary nor Gaussian, coincident detection and other signal consistency checks are required for astrophysical trigger selection and GW detection.

\subsection{Coincidence and vetoes}
\label{coinc}

A GW signal from a given CBC should match with the same template in the common template bank if the SNR is sufficiently high to overcome noise effects. This criterion forms the basis of coincident detection. In the current searches with the
two aLIGO detectors, a coincident trigger must satisfy the following: (i) there are corresponding triggers in each detector - the SNRs must cross the preset thresholds, (ii) the intrinsic parameters recovered independently for each detector are such that they match (the same template clicks) and (iii) the difference in the estimated times of coalescence should not be more than light (GW) travel time between the two  detectors. This difference is allowed a small margin of error because the noise can throw the triggers a little away from their true coalescence times.

To further reduce the false alarms, $\chi^2$ dependent vetoes are applied in the form of newSNR~\cite{Capano:2016dsf,Allen:2005fk} to triggers from each of the detectors. These collected individual detector triggers along with the coincident newSNR statistics are used to estimate the noise background and to assign the statistical significance to the detected GW triggers. We escalate a candidate trigger to a detection, if it shows more than $5 \sigma$ significance.

\section{Hierarchical search}
\label{hierarchical search}

The idea of a two stage hierarchical search is fairly straightforward. First we search over the parameter space by using a coarse grid with a lower threshold on SNR or the detection statistic. The candidate triggers from the first stage are then followed up by finely sampled the parameter space around the neighbourhood (nhbd) of each trigger. The goal is to effectively reduce the number of matched filter computations needed to find a GW signal if it is present in the data. This may also help in reducing the  background arising due to false alarms caused by noise artefacts. The speed-up one gets depends on the coarseness of the first stage bank and the false alarms rate which is related to the choice of first stage signal-to-noise-ratio (SNR) threshold. This procedure is optimised by adjusting the first stage threshold to yield minimum computational cost for the fixed search sensitivity usually defined in terms of sensitivity distance or volume for the CBC searches~\cite{Usman:2015kfa}.

In principle, one could also increase the number of stages of hierarchy, though so far we have restricted ourselves only to two stages.

\subsection{Review of the non-spinning hierarchical search}
\label{hierar_non_spin}

It has been shown previously that a two stage hierarchical search algorithm can be used to speed-up the non-spinning CBC searches by few of orders of magnitude in simulated initial LIGO (iLIGO) like data~\cite{Sengupta:2003wk} and by factor of 7 - 8 in real data from the  second science run (S2) of iLIGO.
The first such study was carried out by \citet{Mohanty:1996bw}. They used only Newtonian waveforms and the hierarchy was performed over just one parameter, namely, the chirp mass. This work was extended to hierarchy over both the masses for 1.5 PN inspiral waveforms by \citet{Mohanty:1996bw,Mohanty:1997eu}. This was then followed up by \citet{Sengupta:2003wk,Sengupta:2001gw} which further extended the  hierarchy to three parameters, namely, the masses and time of coalescence. To incorporate the hierarchy in time of coalescence the data was down sampled in the coarser first stage. 2PN post-Newtonian inspiral-only waveforms were used in their analysis. This most recent work used a geometric template bank placement~\cite{Sengupta:2001gw,Owen:1998dk}. The full details of the previous hierarchical searches with non-spinning GW signal waveforms over simulated and initial LIGO second science run (S2) data are given in~\cite{Sengupta:2004}.


In the latest two stage hierarchical search proposed in \cite{Sengupta:2001gw}, chirp times $\tau_0$ and $\tau_3$ were used instead of individual component masses to create fine and coarse template banks. The template space metric $g_{ab}$ depends very weakly on the chirp times in the parameter space considered. The geometric fine bank with mismatch less than $3\%$ was created using 2PN inspiral-only metric using hexagonal closed packing template placement scheme with iLIGO noise PSD for masses in the range of $(1, 30)M_{\odot}$. In the first stage of the search the data were sampled at a lower rate of $512$ Hz and the coarse template bank was created with mismatch less than $20\%$, that is, MM of $0.8$. For such large values of mismatch, the metric approximation breaks down. Therefore, the coarse bank is created numerically by a rectangular placement of the templates along the $\tau_0$ axis.
In the first stage, the lower MM reduces the number of templates in the bank significantly. Moreover, downsampling reduces the cost of each FFT in each MF operation. However, this reduction in computational cost comes at the cost of reduced SNR of the recoverable signal. Hence, in order to ensure that we do not lose an otherwise detectable GW signal, the applied SNR threshold must be lower than the one used in the single stage flat search which is the usual search with the bank of MM $> 0.97$.
With the individual detector SNR thresholds of $6$ and $8$ for the first and second stage respectively, the search showed computational cost reduction by few orders of magnitude for simulated data with Gaussian noise~\cite{Sengupta:2001gw,Sengupta:2003wk} and almost by an order of magnitude during search with iLIGO S2 data.

All the earlier works mentioned above considered only a single detector and did not use any signal consistency tests such as the $\chi^2$ discriminator. Apart from introducing those essential components in the search to make the implementation applicable for real data, there are two primary routes to further extend the hierarchical search strategies, either by increasing the number of stages in the hierarchy or by including more parameters in the two-stage hierarchy or both. Since the current CBC waveforms include spins, we have opted for the latter. We may explore the feasibility of the former option in future.


\subsection{Hierarchical search with aligned-spin waveforms}
\label{hierar_aligned_spin}

In this work we explore the possibility of a hierarchical algorithm for CBC searches with non-precessing template waveforms in the modern set up. We  use the {\tt PyCBC} pipeline~\cite{Usman:2015kfa} with LIGO's O1 type of search ~\cite{Capano:2016dsf}. We use the full inspiral-merger-ringdown (IMR) aligned spin waveforms with dominant $(2, ~2)$ mode. Both coarse and fine template banks are generated using stochastic template placement algorithm~\cite{Harry:2009ea}. We also introduce signal consistency vetoes and two-detector coincidence in the search, which were not part of previous efforts.

\subsubsection{Proof of concept: Zero-noise case}
\label{proof_of_concept}


Before moving on to the full-fledged pipeline, we present our initial study with zero-noise BBH injection case. We use AdvLIGO PSD for computing the match using inner product described in Eq.~(\ref{eq:mf}). This also motivates our choices for thresholds and size of the fine bank nhbds for each of the coarse bank trigger template.

We choose 2000 binary black hole (BBH) injections in H1-L1 detectors with single detector optimal matched filter SNR in the range of 5 to 15 for each of the detectors. Both the BHs have masses uniformly sampled from the range of (5, 10) $M_\odot$. Both the BHs can have spin components along the orbital angular momentum in the range of (-0.98, 0.98). The injections are uniformly spread all over the sky. For this study, we use the actual matched filter SNRs and coincident SNRs without $\chi^2$ (the $\chi^2$ weighed SNRs are not applicable here).

\begin{figure}[h]
\begin{center}
\includegraphics[width=0.5\textwidth]{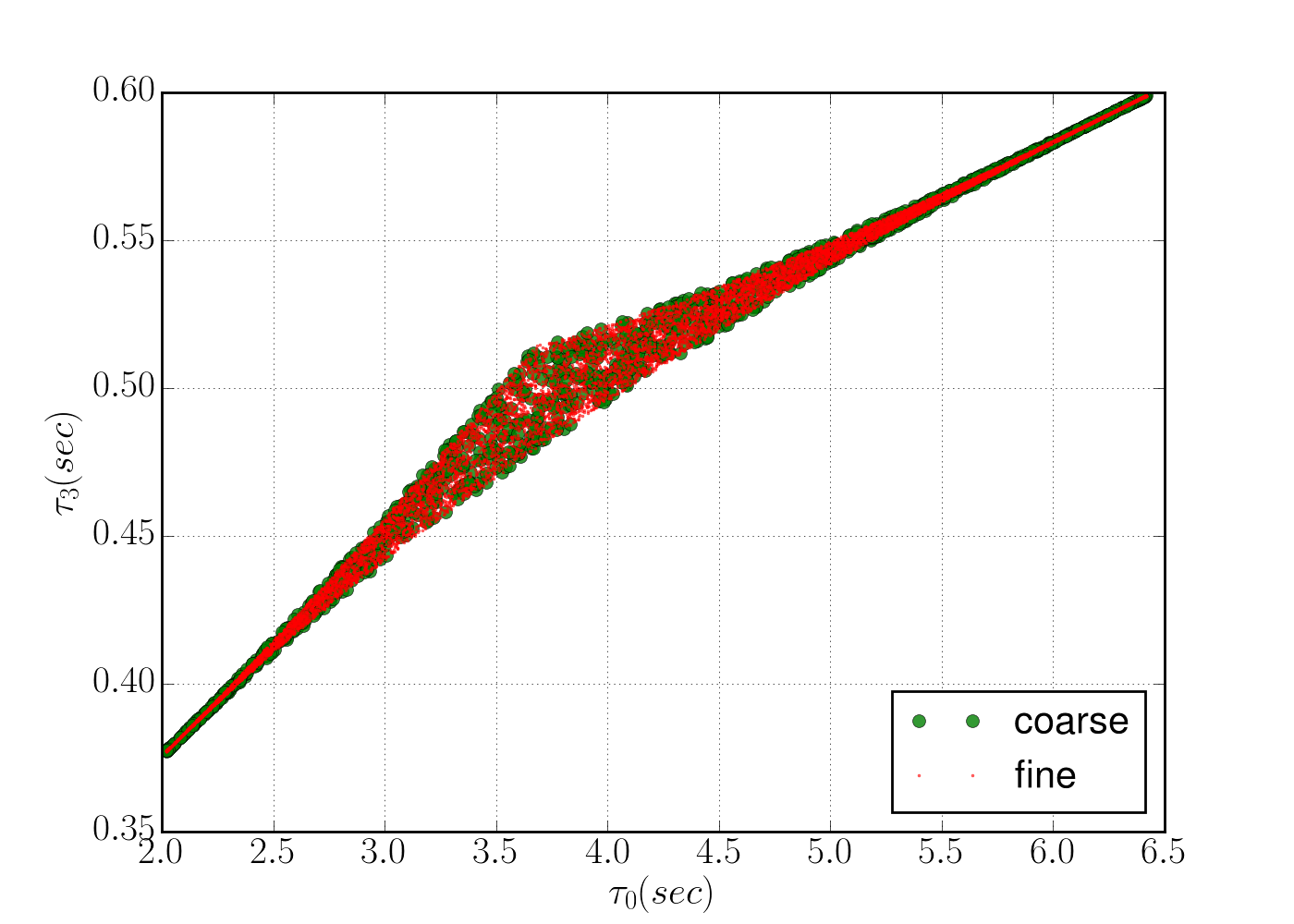}
\includegraphics[width=0.5\textwidth]{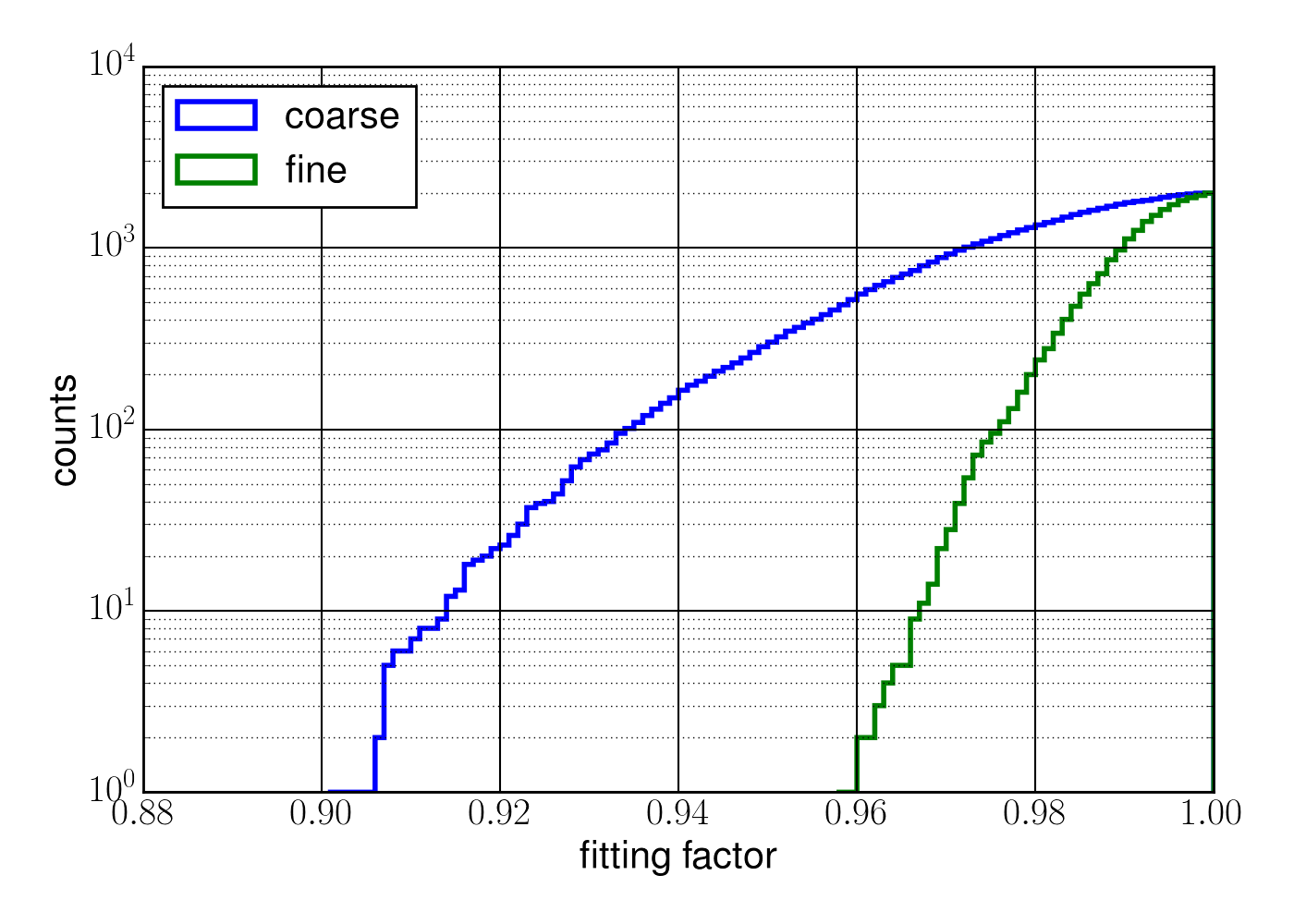}
\end{center}
\caption{The figure on the top shows the non-spinning template banks in $\tau_0-\tau_3$ plane and the one on the bottom shows that the bank do no have any holes as all the fitting factor (FF) values are greater than the MM values used to construct each of the banks}
\label{zero_noise_bank}
\end{figure}

We construct a coarse and a fine template bank with MM of 0.9 and 0.97 respectively for the same parameter space in component masses and spins as used for the BBH injections. Both the banks are created using template space metric as described in Section~\ref{bank}. This is possible because we have used {\tt TaylorF2RedSpin} approximant in this study for which the analytic metric is available. The coarse bank has $1200+$ templates while the fine bank contains $10000+$ templates, that is, the fine bank is about eight times denser than the coarse bank. The banks and the proof that there are no holes (i.e. the prescribed MM condition is satisfied in both the cases) are shown in figure~\ref{zero_noise_bank}. We choose sampling rate of 512 Hz for the construction of the coarse bank, which is used for the Stage-I in the search. Then we use sampling rate of 1024 Hz for the stage-II with the fine bank. We deliberately choose these sampling rates for stages I and II of the zero-noise study as ISCO frequencies for BHs having masses in the range 10-20 $M_\odot$ are in range 220-440 Hz. With the sampling rate of 512 Hz and 1024 Hz, we have the Nyquist frequency of  256 Hz and 512 Hz respectively. Reduction in sampling frequency leads to loss of SNR for some of the templates in the coarse stage as compared to the generic flat search. The template duration for all the signals in the consideration is less than 8 sec. Hence the data segments length of duration 16 sec has been chosen for computing matched filters.

\begin{figure}[h]
\begin{center}
\includegraphics[width=0.5\textwidth]{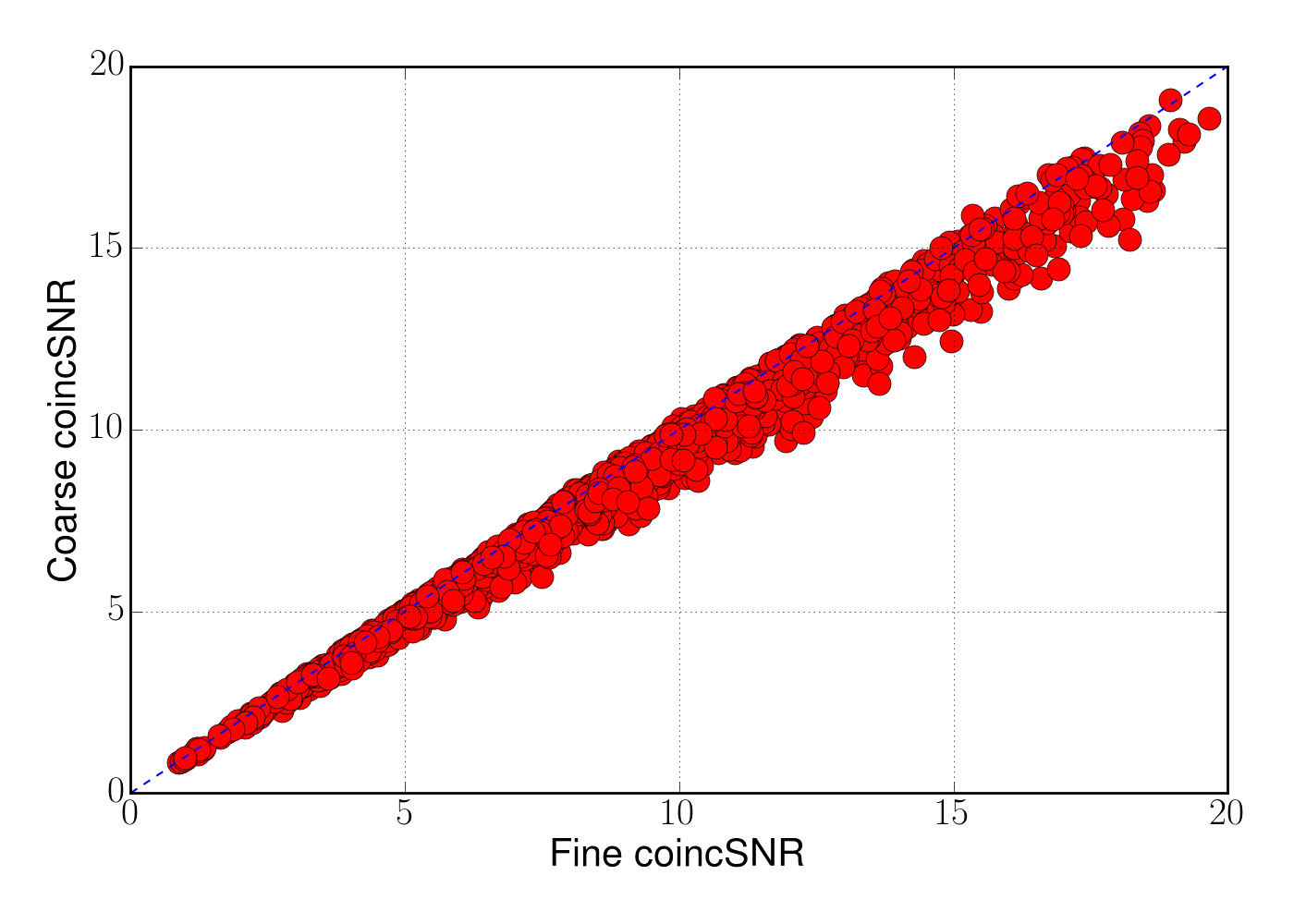}
\end{center}
\caption{Red circles show the coincident SNR for the fine and coarse searches. It shows how much SNR is lost in stage I with the lower sampling rate and the coarse bank which stage II should recover with fine subbank and full sampling rate.}
\label{zero_noise_coinc_fine_coarse}
\end{figure}

We then perform matched filtering using stage I and stage II banks for each of the injections for both H1 and L1 detectors and compute coincident SNRs. We compare the SNR in Stage-I with the SNR in stage II in Figure~\ref{zero_noise_coinc_fine_coarse}. The loss of SNR is due to the coarse sampling of the parameter space and reduced sampling frequency as mentioned above. Then, in Figure~\ref{zero_noise_max_snr_loss}, we plot the maximum possible loss of SNR along the horizontal axis and in the vertical axis the match between the templates that correspond to maximum SNR (trigger templates) in stage I \& II. The match between stage I and stage II trigger templates corresponding to the same injection tells us how large of a nhbd is required if we want to recover full SNR in stage II of the hierarchical search. The figure tells us that even for zero noise case, we have to consider nhbd which is large enough to incorporate templates corresponding to as low as 85\% match in order to recover the full SNR in stage II of the hierarchy. With noisy data, this number will be lower. Hence, in our 2-stage hierarchical search, we construct the fine bank in the nhbd for stage I trigger templates by combining all the templates in the fine bank having match greater than 75\% with the same coarse bank template. Figure~\ref{zero_noise_max_snr_loss} also shows that maximum SNR loss is $85 \%$ in stage I. This tells us how much lower we should keep our stage I SNR thresholds, for individual detectors ($\rho_{\textrm{single, I}}$ and for coincidence $\rho_{\textrm{coinc, I}}$) as compared to flat search. We keep stage I threshold at about 90 \% SNR loss as discussed later in~\ref{Formalism}.

\begin{figure}[h]
\begin{center}
\includegraphics[width=0.5\textwidth]{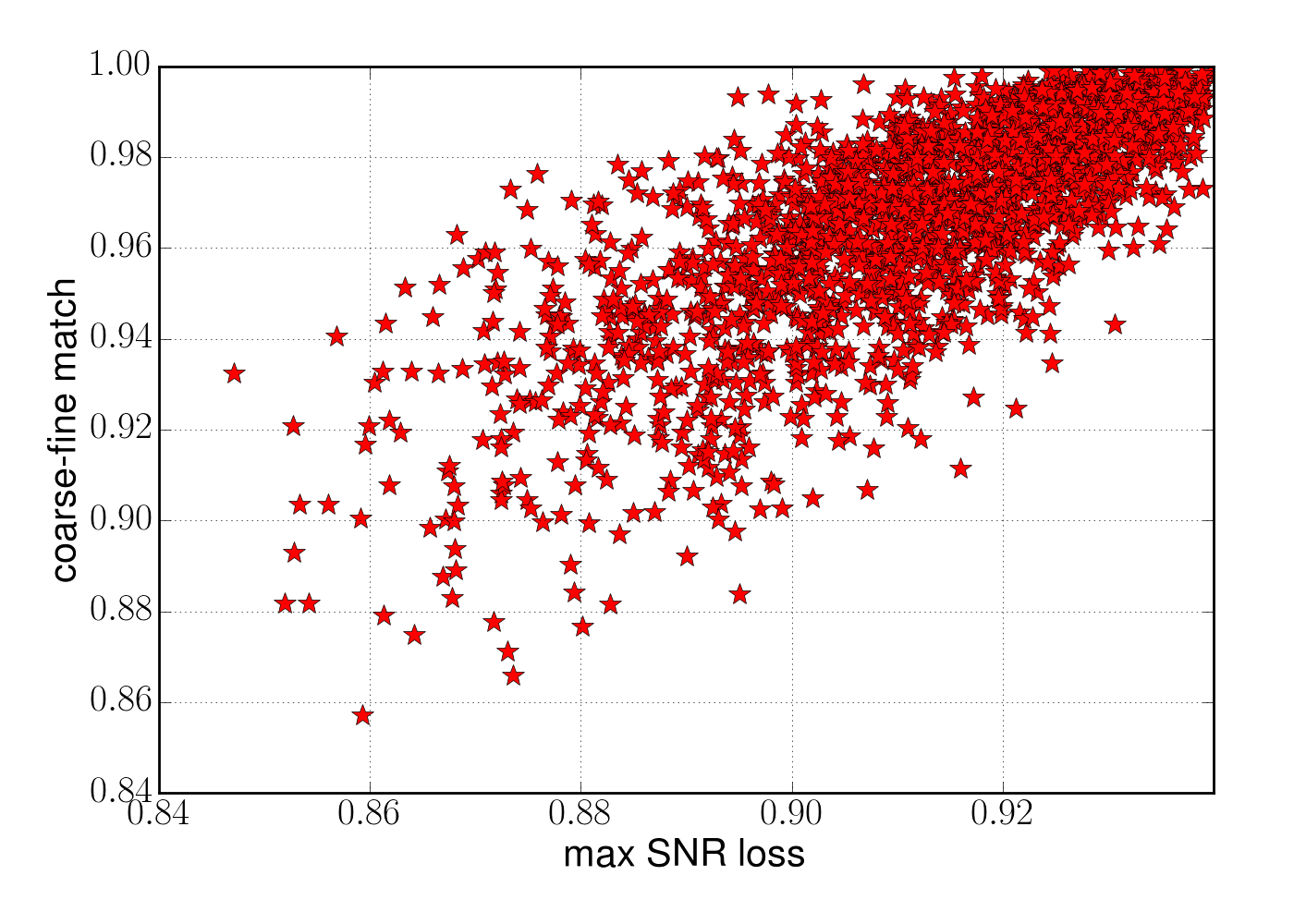}
\end{center}
\caption{In the figure, Each red star represents maximum loss of the SNR (normalized) in stage I against the match between the stage I (coarse) trigger template and stage II (fine) trigger template for a given injection.}
\label{zero_noise_max_snr_loss}
\end{figure}

Figure~\ref{zero_noise_coarse_hl} shows the recovered stage I SNRs for injections in each of the detectors H1 and L1. The figure shows us variation in the SNR across the detectors just because of their response to the same signal depending upon the signals sky location which is encoded into the antenna pattern functions. The variation in the single detector SNRs gives the idea about the signals which the current setup of the 2-stage hierarchical search may not detect which flat search may. If the signal is barely above the single detector SNR ($\rho_{\textrm{single, flat}}$) in one of the detectors, then, even after reducing the single detetor SNR threshold ($\rho_{\textrm{single, I}}$) more than 90\% compared to the $\rho_{\textrm{flat, I}}$, we may miss it as the noise may not allow the correct template to come out as stage I trigger template for the signal in at least one of the detectors. Due to this, stage II would not look for that signal even though it's loud enough in the other detector. We think these signals, most likely, will not be strong enough to stand against the overall noise background with a much of a significance. But there is a chance that the hierarchical search may miss some very few signals having pretty low SNR in one of the detectors. However, with the increase in the number of detectors and with their improved duty cycles, this is unlikely to happen.

\begin{figure}[h]
\begin{center}
\includegraphics[width=0.5\textwidth]{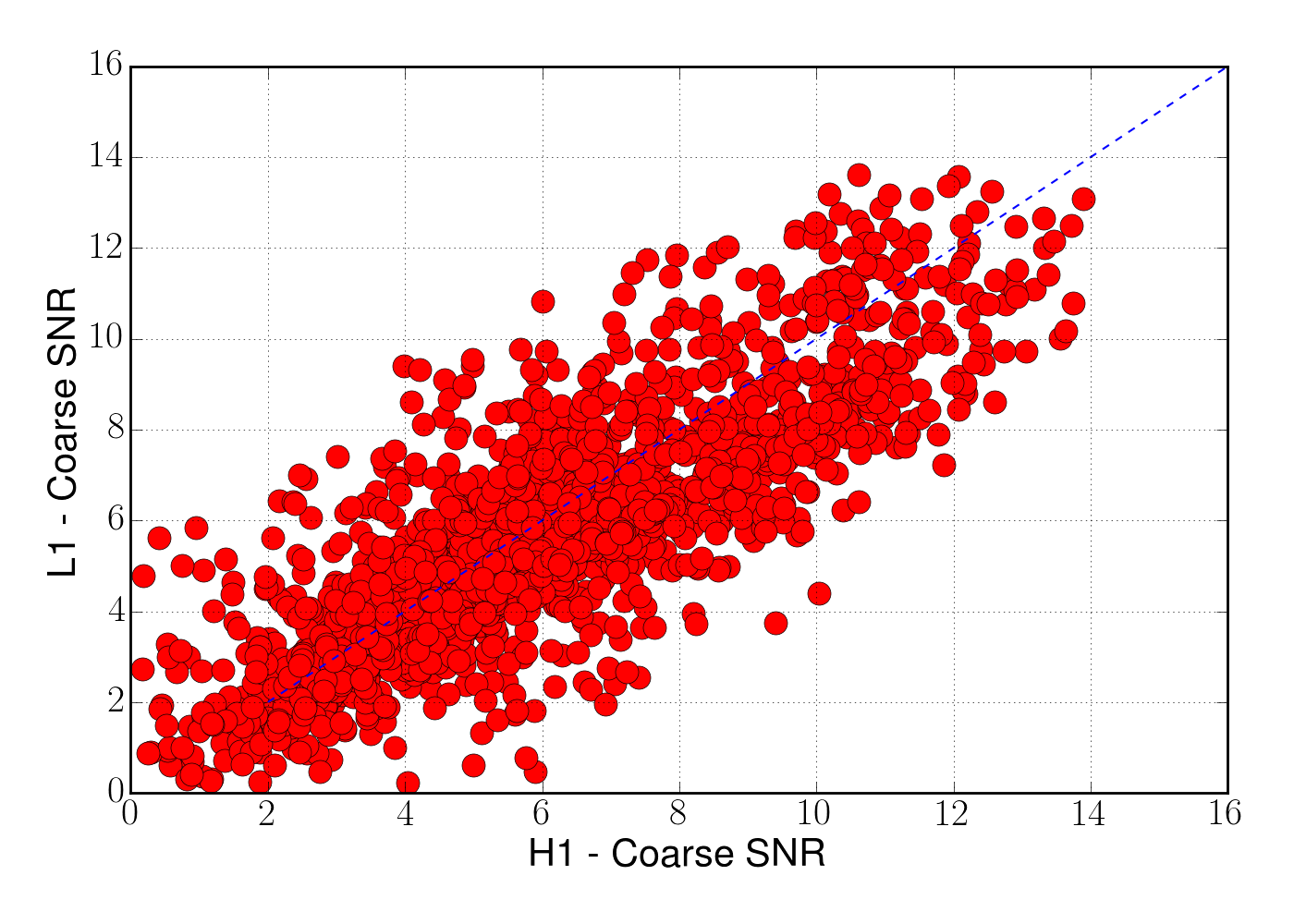}
\end{center}
\caption{The figure shows the recovered SNR for each of the injection as observed in detectors H1 and L1. The large variation in the recovered SNR for a given injection across the detectors is due to antenna pattern functions. The dashed line represents equal SNR in each of the detectors.}
\label{zero_noise_coarse_hl}
\end{figure}

To demonstrate the point mentioned above, we plot missed-found triggers in figure~\ref{zero_noise_miss_found} for the zero noise study. We have chosen $\rho_{\textrm{single, flat}}=5$ and $\rho_{\textrm{coinc, flat}}=8$ for the flat search. If the recovered coincident SNR of the injection is greater than $\rho_{\textrm{coinc, flat}}$ then we claim it as a detection in the flat search. For the 2-stage hierarchical search, we choose $\rho_{\textrm{single, I}}=4.5$ for both the detectors and $\rho_{\textrm{coinc, I}}=7.2$ for stage I. We follow the trigger in the second stage only if its recovered stage I coincident SNR is more than $\rho_{\textrm{coinc, I}}$. Then to claim the detection after stage II, we follow the same thresholds as for the flat search i. e. $\rho_{\textrm{single, II}}=5$ and $\rho_{\textrm{coinc, II}}=8$. Then we claim detection for the hierarchical search after stage II. To imitate the behaviour of stationary Gaussian noise with the coincident matched filter statistics, we have used some fiducial SNR cut-offs. We assume stage I will miss the injected signal if the recovered single detector SNR corresponding to the injection is less than $4.5$ in one and less than $6.5$ in the other detector. For stage II and the flat search, the single detector SNR  should not be less than $5$ in one and less than $6.5$ in the other simultaneously. We assume the signal will be missed in the stage I or the flat search if recovered SNRs are below the above mentioned values as noise may give coincident triggers from completely random template. Then the hierarchical search may not be able to follow-up the trigger with the correct fine bank in the nhbd. With these numbers, we have shown found-missed triggers with stage I (coarse) coincident SNR and flat (fine) coincident SNR in figure~\ref{zero_noise_miss_found}. We found that there are two injections which the hierarchical search may miss or recover with lower coincident SNR which the flat search is likely to detect. They are shown by blue triangles. The greens are the injections which both the searches found and the red ones are missed by both the searches. The region near blue triangles in the region where the current hierarchical search may miss the signal or most likely to detect them with slightly lower significance as recovered SNR may be lower than the flat search. With these caveats in mind, we describe the whole setup of the current 2-stage hierarchical search in the following subsections.

\begin{figure}[h]
\begin{center}
\includegraphics[width=0.5\textwidth]{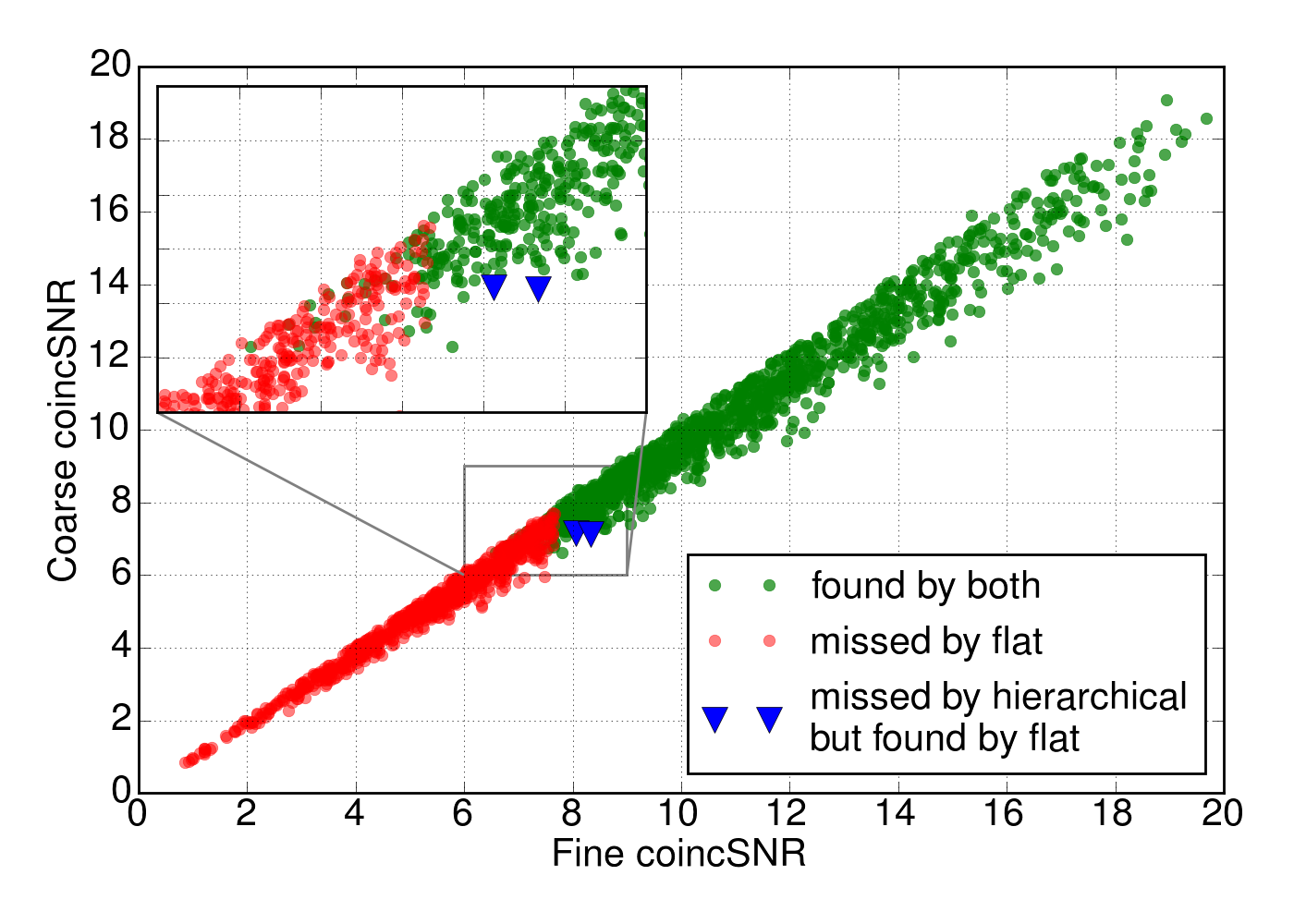}
\end{center}
\caption{Injections found and missed in the coarse and fine searches}
\label{zero_noise_miss_found}
\end{figure}

\subsubsection{Formalism}
\label{Formalism}
Now we describe our current set up of the 2-stage hierarchical search pipeline where we introduce coincident detection for two detectors. The full search is illustrated by the flowchart in figure~\ref{hflow}. We start by creating stochastic coarse and fine banks for the intrinsic parameters (these are masses and aligned spins) which have MM of $0.9$ and $0.97$ respectively. For template banks we use the harmonic PSD which is the harmonic mean of PSDs of H1 and L1 detectors during O1. We use the same PSD to generate simulated Gaussian noise for both the H1 and L1 detectors. These data are then divided into smaller chunks of 4096 sec each for estimating the local PSD which is required for the matched filtering computations. The matched filtering is done with data segments of duration 256 sec and with 128 sec overlap with the previous segment. This overlap is needed because we must discard data from both the ends of a data segment due to the circularity property of the FFT algorithm and also get rid of other numerical artefacts~\cite{Usman:2015kfa,Brown:2004vh}. We therefore actually search only 128 sec of data in one matched filter computation.

\begin{figure*}
\begin{center}
\includegraphics[width=\textwidth]{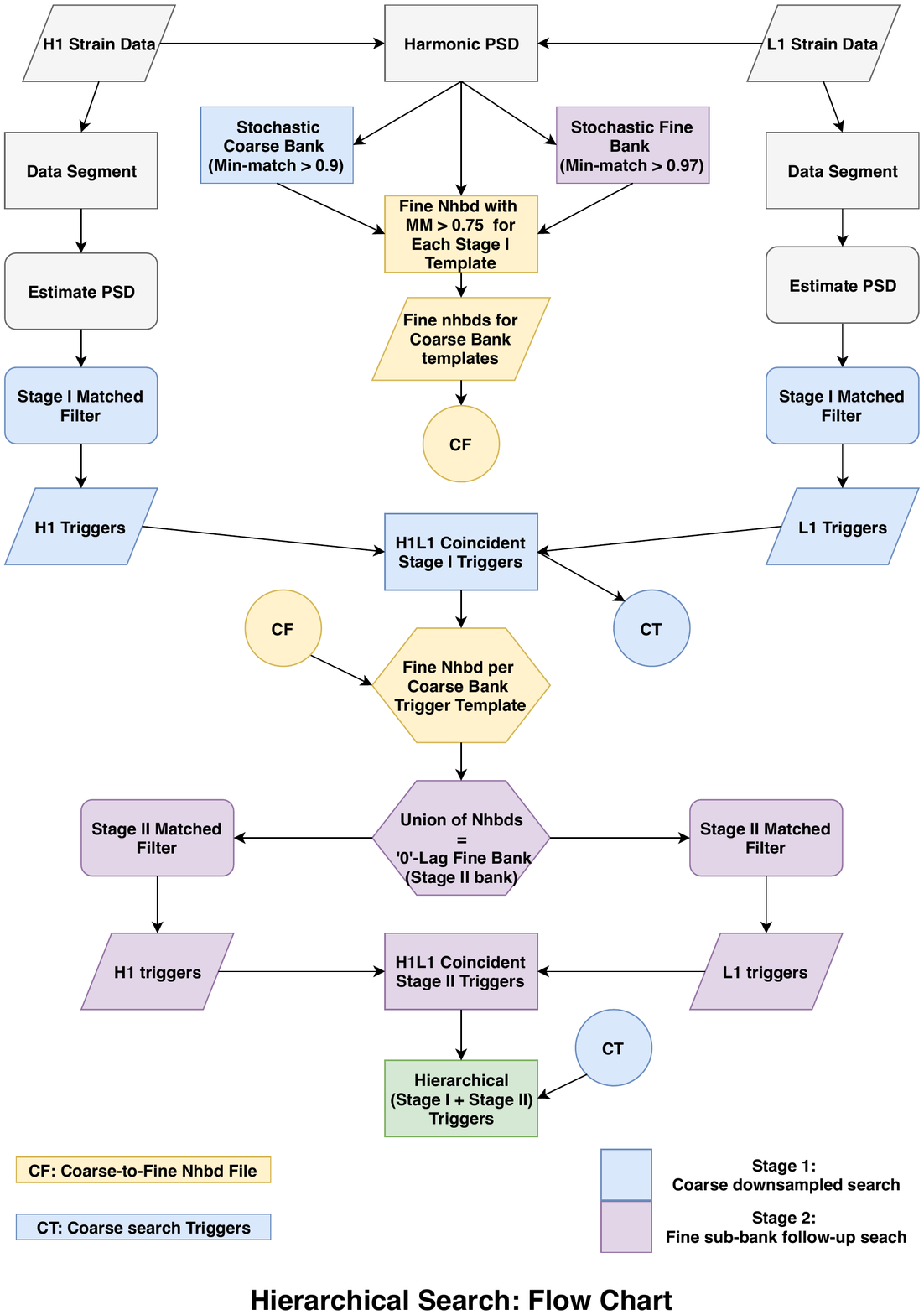}
\end{center}
 \caption{Flow chart describing the set up of the current 2-stage hierarchical search}
\label{hflow}
\end{figure*}

The hierarchical search begins with the first stage,
where data are sampled at a lower rate and with a coarse bank for each detector. Single detector events are recorded if the statistic crosses a pre-determined threshold for each  detector. The statistic employed here is the power $\chi^2$ re-weighted new-SNR~\cite{Usman:2015kfa,Capano:2016dsf}. In the first stage the threshold is lower than the second stage threshold. We then compare parameters of the triggers from each detector and select only those triggers whose parameters match - these are the coincident triggers for the first stage. We then follow up these candidate triggers with a fine search using a fine sub-bank constructed around each coincident trigger.

To create the fine sub-bank from the zero-lag coarse coincident triggers, which would be common for all the detectors, we combine the precomputed fine bank nhbd for each of the coarse trigger template. Then we repeat the search with a higher sampling rate and with time segment specific fine sub-bank and collect the individual detector triggers from each of the detectors H1 and L1. We collect coincident triggers from the second stage single detector triggers.
Then we combine these zero-lag coincident triggers from both the stages to give us the foreground triggers.

The noise background is obtained as follows: We use combined (coarse + fine sub-bank) individual detector triggers in the time-slides to get the hierarchical search noise background. This background is usually lower than the flat search (single stage search with the fine bank~\cite{Usman:2015kfa}) background as it comes from significantly reduced number of templates. The fine sub-bank contributes negligibly to the time slide background because it has much fewer templates which can be common in the time slides due to its construction using zero-lag triggers only. This is where our 2-stage hierarchical search deviates from the flat single stage time slide background.

In principle, this hierarchical background can be used to assign significance to the foreground triggers using scaling as described later. But we use only zero-lag sub-bank in the second stage, it is different from the current search background estimation and may show some bias. Hence we have used the usual flat search background to estimate the search sensitivity distance in the current study. With the hierarchical background, the sensitive distance for the hierarchical search will slightly improve than what has been shown section~\ref{hierarchical_result}. Further study is required to improve the estimate of the background and see if there is a generic way to obtain an equivalent robust background.

We can of course reproduce a background for the hierarchical search analogous  to the flat search. We must then construct non-zero time lag fine sub-banks for the coarse stage triggers. But then these sub banks together will cover most of the fine bank, thus losing any computational advantage that we may have obtained.

\subsubsection{Parameters used in the hierarchical search}
\label{first_stage}

In this subsection we provide the detailed description of the parameters used in our hierarchical search. We consider the same ranges of masses and spin parameters as were employed in the search during the first aLIGO science run (O1). We also use the same waveforms as those employed in O1. For aligned-spin GW signals, our set of intrinsic parameters are component masses and spins along orbital angular momentum. For the individual masses, the heavier mass is in the range $(1, 100)M_\odot$ and secondary mass is in the range $(1, 50)M_\odot$. In case of neutron stars we have taken the  masses to be in the range $(1, 3)M_\odot$ and spin components along the orbital angular momentum from $-0.4$ to $0.4$. Black holes have spin component values ranging from $-0.9895$ to $0.9895$. The fine and coarse banks are generated using stochastic template placement algorithms (ueberbanks) with mismatch of $3\%$ and $10\%$ respectively. The choice of $10\%$ mismatch for the coarse bank is somewhat arbitrary. With these numbers and with O1 harmonic power spectral density (PSD), we obtained $\sim~60000$ templates for the coarse bank and $\sim~250,000$ templates for the fine bank. One observes that the fine bank is almost $4$ times larger in comparison with the coarse bank. For the search templates, we have used {\tt TaylorF2} approximant for total mass less than $4 M_{\odot}$ and {\tt SEOBNRv2\_ROM\_DoubleSpin} for the rest of the parameter space.

Even though the choice of the MM $= 0.9$ may look somewhat arbitrary, we have tried other values of MM, for example, 0.8 (as employed in the previous hierarchical searches) and 0.85. We found that the overall loss in the SNR is unacceptable because of loss in sensitivity. Therefore, we fix MM at 0.9 which gives a coarse bank with about quarter the number templates as compared to the usual fine bank with MM $= 0.97$.

O1 harmonic PSD is used for the template banks and for simulating data with the lower cut-off frequency set at $30$ Hz. For the first stage in the search, we sample the data at a reduced rate of $512$ Hz while for the second stage, we sample the data at $4096$ Hz. Because of the reduced sampling rate of $512$ Hz in the first stage, we must cut off the signal below 256 Hz. However, we recover more than $90\%$ of the signal SNR, for all the $10000$ non-precessing injections. We also ensure that the banks do not have ``holes".

As mentioned earlier, 5 days of simulated coincident data for the two LIGO detectors H1 and L1 are used assuming both of them have the harmonic PSD of O1 run. We inject $>~10000$ non-precessing CBC signals in the data with the parameter ranges as mentioned earlier. Injections were uniform in volume, orbital inclination and coalescence phases. Injections were distributed as follows: $>~2000$ double neutron star (DNS), $>~4000$ neutron star- black hole (NSBH) and $>~4000$ binary black hole (BBH). Neutron star masses were  in the range $(1,~ 3)M_\odot$. Further, the injections were uniformly distributed in the total mass. All the injections were with aligned-spin. The   optimal SNRs for the injections were in the range $(8,~ 30)$. For DNS injections, we have used {\tt TaylorT} approximant for injection and {\tt IMRPhenomD} and {\tt SEOBNRv2} for NSBH and BBH injections respectively.

Apart from the above, we injected more than $8000$ precessing signals with total mass in the range of 5 to 150 $M_{\odot}$ with the dominant mass ranging from 4 to 100 $M_\odot$. For the precessing injections, we have used {\tt IMRPhenomPv2} approximant.

Next we go on to the first stage search and describe the coarse triggers and the fine subbank used in the second stage.

\subsubsection{Stage I triggers}
\label{first_stage}

The goal of the first stage is to obtain candidate triggers which will then be followed up in the second stage of the search. To obtain these, we need to decide a threshold on the detection statistics, which is the  chi-square weighted newSNR~\cite{Capano:2016dsf} for the single detector statistic and coincident newSNR i.e. newSNR of single detectors added in quadrature for the pair of LIGO detectors. This statistic is used for coincident triggers for both the stages and also for the flat search. The subscripts ${\rm I}$ and ${\rm II}$ will refer to the first and second stages of the hierarchical search respectively. We decide on the individual detector thresholds $\rho_{\rm single, flat} = 5.0 = \rho_{\rm single, II}$ where $\rho_{\rm single, flat}$ is the threshold for the flat search and
$\rho_{\rm single, II}$ is the threshold for the second stage search. We  decide to keep single detector newSNR ($\rho_{\rm single, I}$) to be $4.5$ which is $90 \%$ of $\rho_{\rm single, flat}$. We have chosen these values because we expect SNR loss to be less than $10 \%$. The amplitude of the GW signal in frequency domain scales as $f^{-7/6}$ and the SNR in the first stage is reduced both because of the coarse template bank and a lower sampling rate, which we denote by $\rho_{\rm reduced}$. The results are as follows:
\begin{equation}
 \rho_{\rm reduced} ~=~ \MM_{I} \frac{\rho \left(f_l, f_{u, I} \right)} {\rho \left(f_l, f_{u, II} \right)}
\label{snr_loss}
\end{equation}
where we have defined $\rho$ as:
\begin{equation}
 \rho \left(f_l, f_u \right) = \int_{f_l}^{f_u} \frac{f^{-7/3} }{S_n \left( f \right)} ~df
\end{equation}

For the values we have chosen, we get $\rho_{\rm reduced}$ $>$ 88 \% for $\MM_{I}$ of 0.9 for the coarse bank. But if we use the factor $\MM_{I} / \MM_{II}$ instead of $\MM_{I}$ in equation~\ref{snr_loss}, we get $\rho_{\rm reduced}$ $>$ 91 \%.

\begin{figure}[h]
\begin{center}
\includegraphics[width=0.5\textwidth]{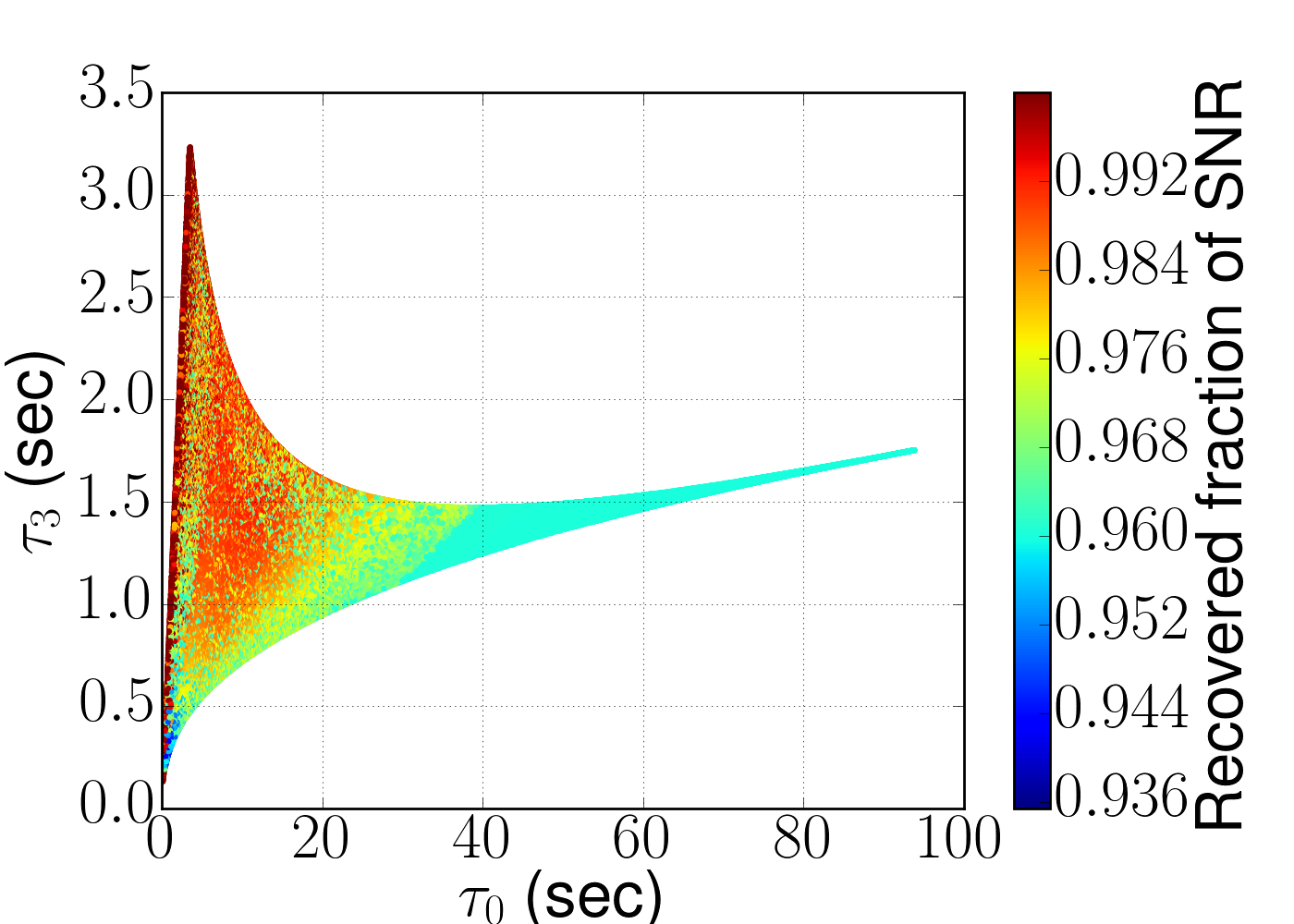}
\caption{Reduced SNR due to lowered sampling rate over the parameter space in consideration. The maximum SNR loss is $\sim~6 \%$.}
\label{low_sampling_snr_loss}
\end{center}
\end{figure}

For obtaining triggers for the first stage, data segments of length $256$ sec are taken sampled at the rate $512$ Hz from each of the H1 and L1 detectors. Then the procedure as described in subsection \ref{Formalism} is followed.

\subsubsection{Stage II search}

After obtaining the coincident triggers obtained in stage ${\rm I}$, we proceed to the stage ${\rm II}$. Here we construct a fine bank in a small neighbourhood around each stage I trigger. This is the fine subbank associated with the trigger. The neighbourhood for the subbank is so chosen that the templates in the fine bank have a match more than $0.75$ with the trigger template from the coarse bank. We have chosen a much smaller value for the match in order to compensate for the noise effects and other factors. In this way, for each of the coarse bank template we have an associated  neighbourhood and a fine subbank. These fine subbanks are pre-calculated for each coarse template.

\begin{figure}[h]
\begin{center}
\includegraphics[width=0.5\textwidth]{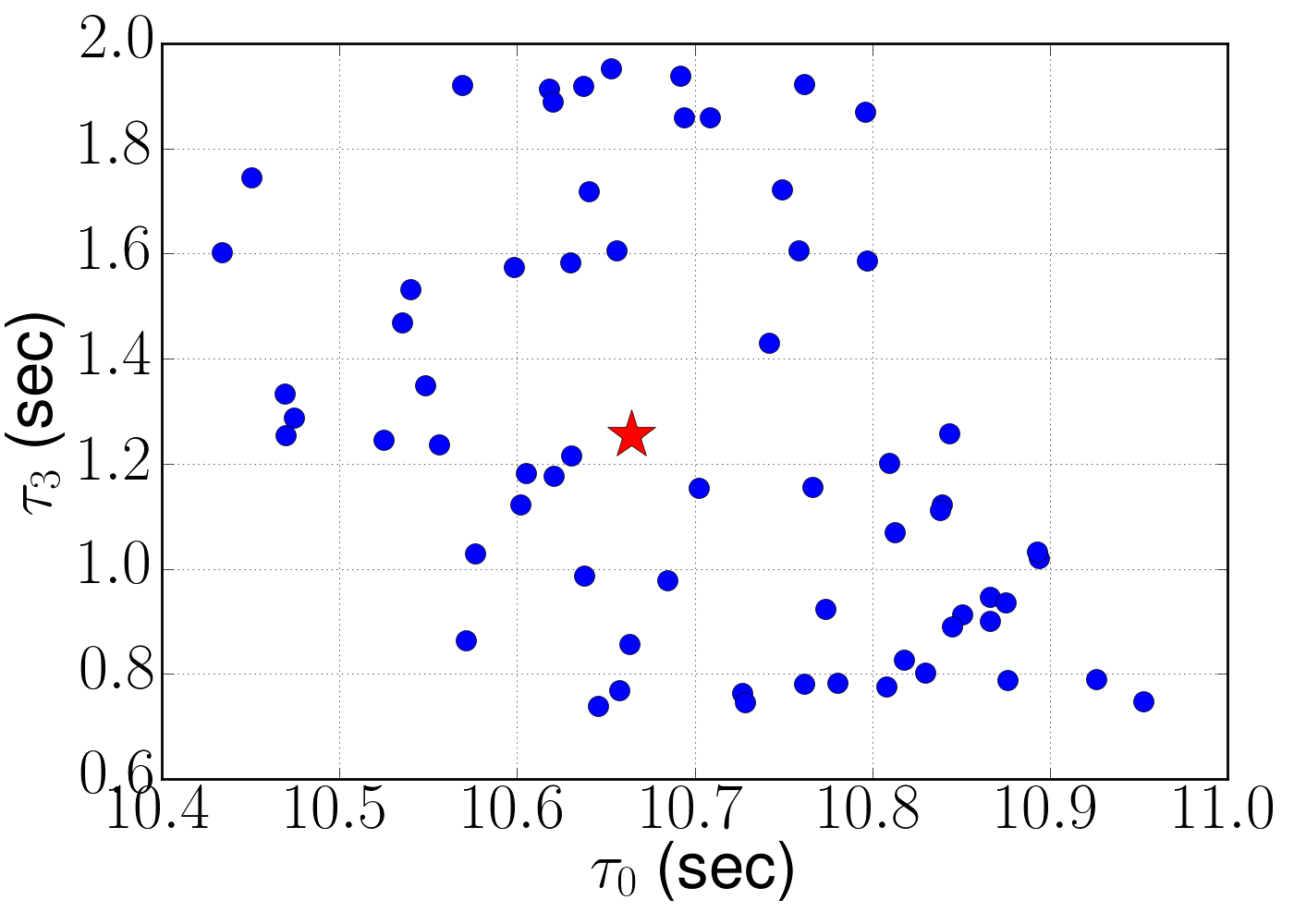}
\includegraphics[width=0.5\textwidth]{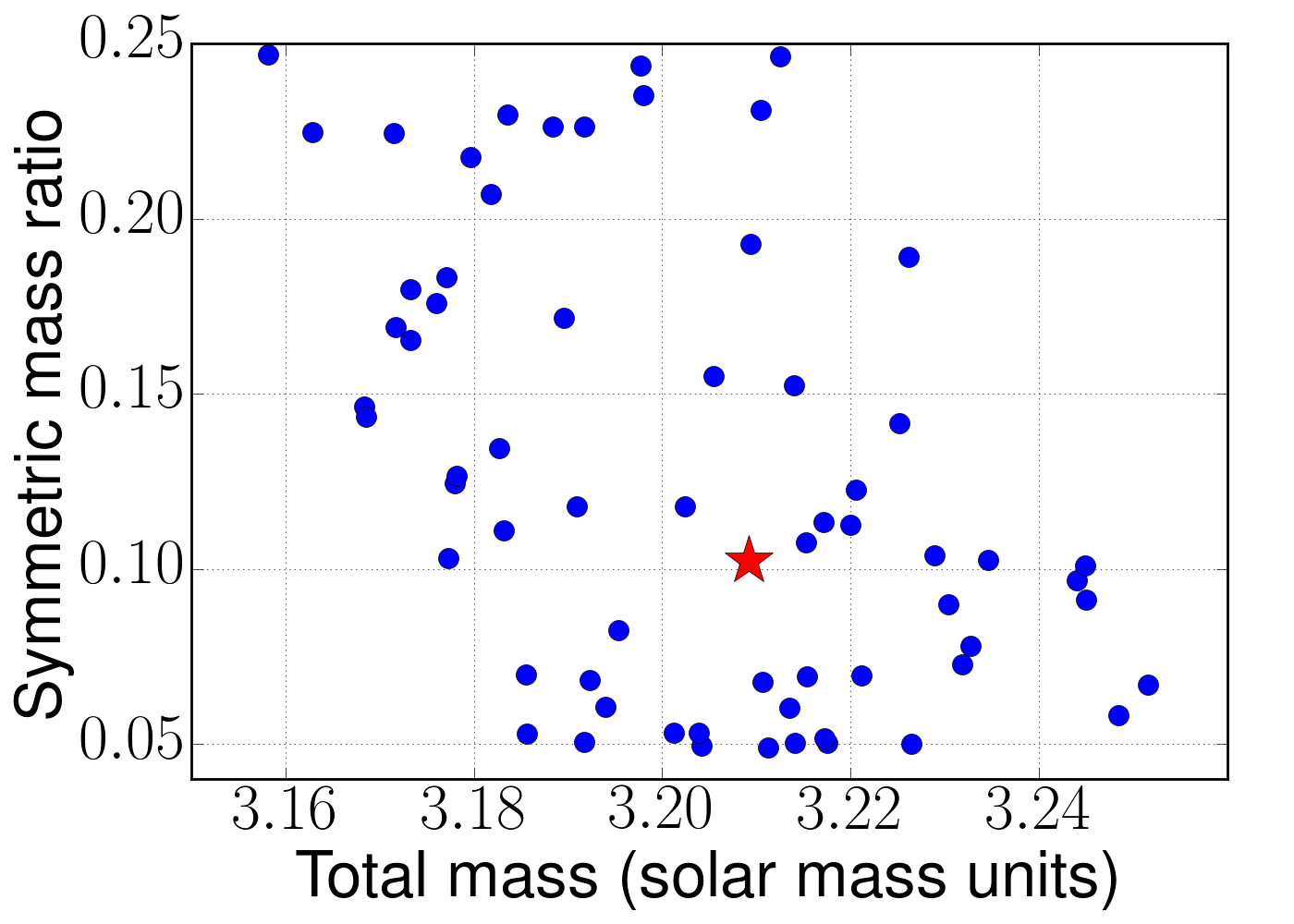}
\caption{A typical fine-bank neighbourhood of a coarse template projected onto mass parameters. A red star shows the coarse bank template and blue dots are the templates in the fine bank with FF values greater than 75 \%. There are 65 templates in this fine-bank neighbourhood. }
\label{fine_bank_nhbd}
\end{center}
\end{figure}

\begin{figure*}
\begin{center}
\includegraphics[width=0.475\textwidth]{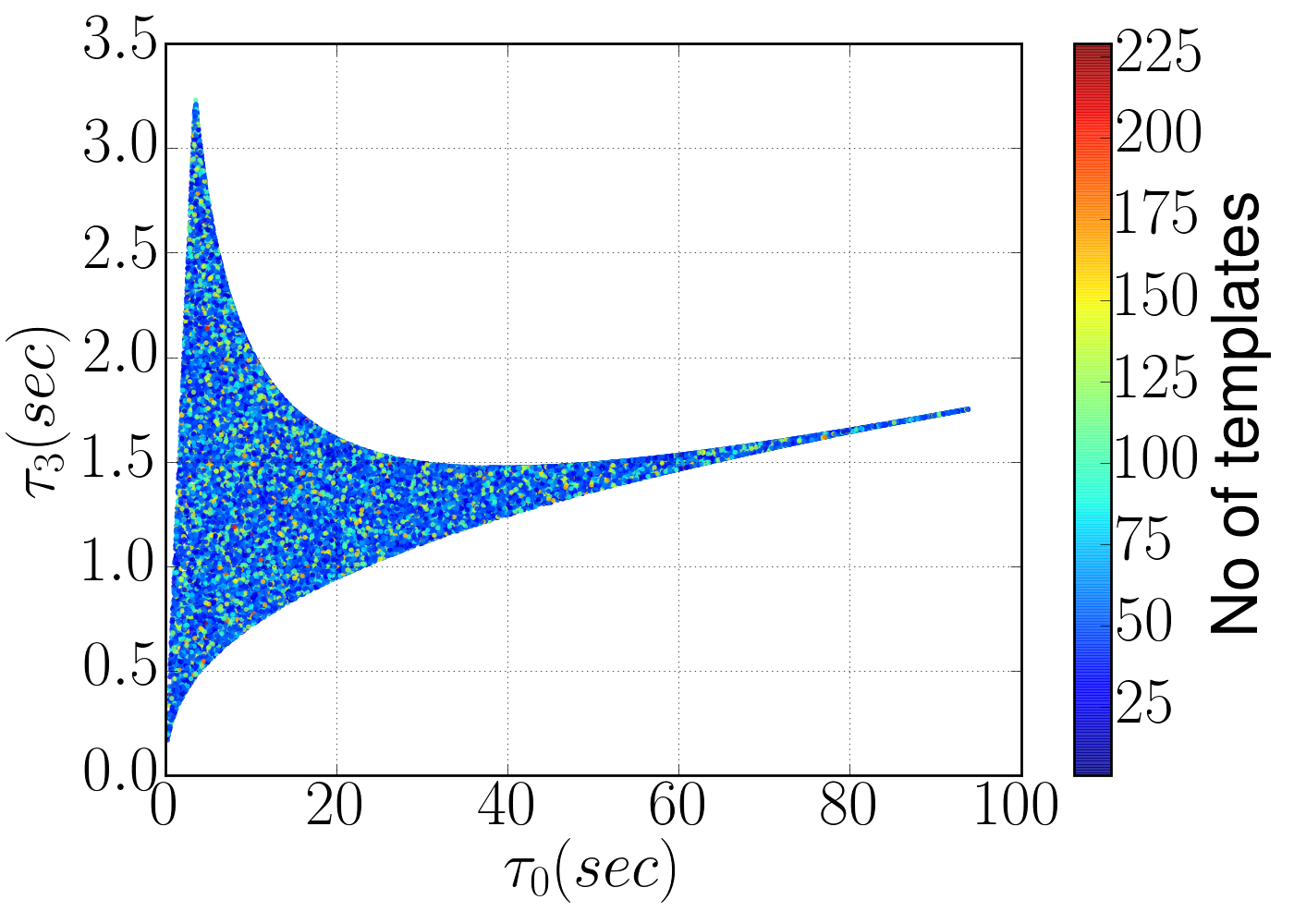}
\includegraphics[width=0.475\textwidth]{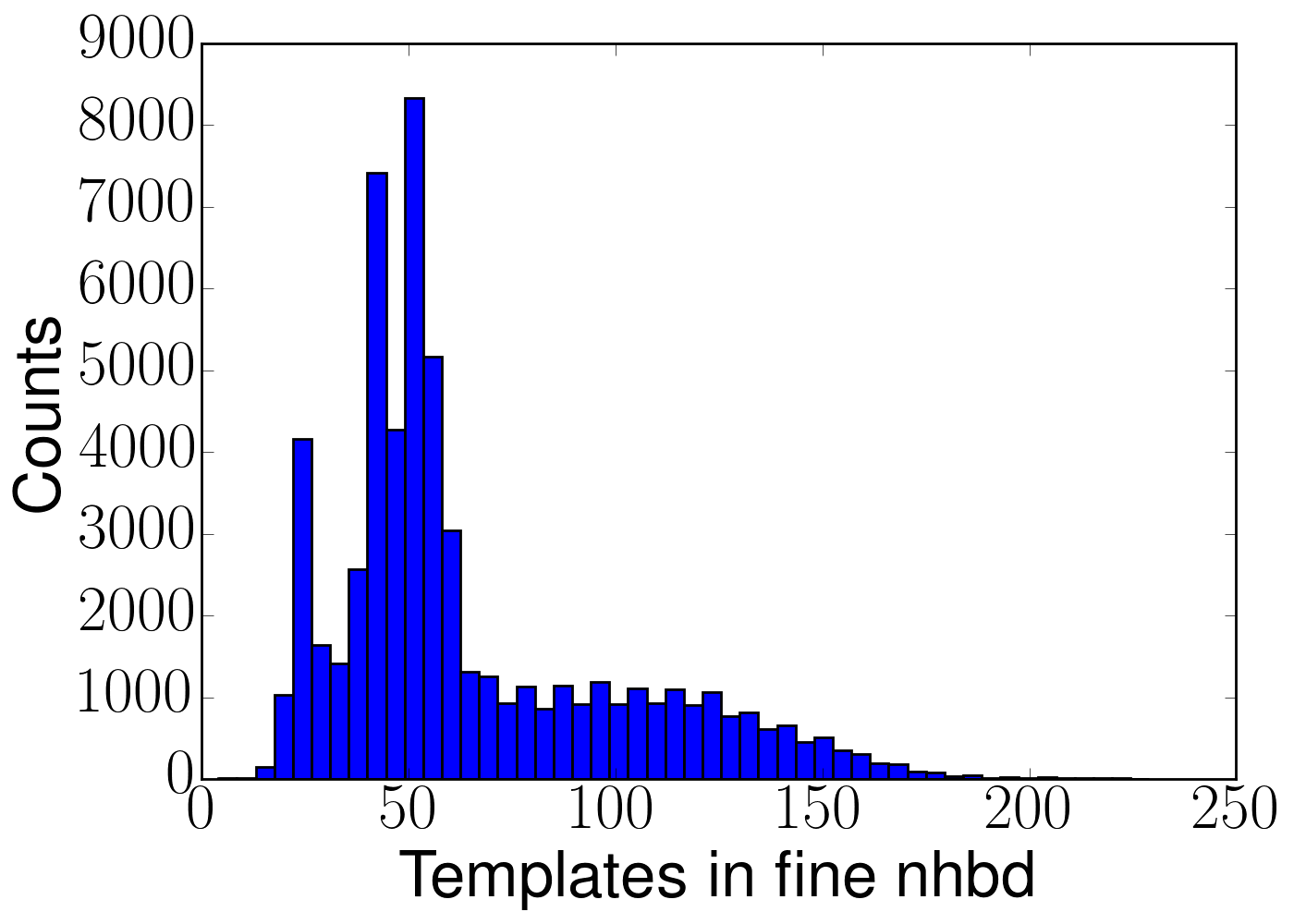}
\caption{Number of templates in the fine-bank neighbourhood for the coarse trigger template. Left plot denotes number of templates in the fine nhbd projected over chirp times. On the right, we have a histogram showing the count of coarse templates having different number of templates in the respective fine nhbd. This helps in the understanding the size of the fine sub-bank for any generic data segment.}
\end{center}
\end{figure*}

We obtain coincident coarse triggers for each data segment and then we proceed to create a fine subbank corresponding to that segment. Then we take the union over all the data segments of all the fine subbanks. This unified subbank depends upon single detector statistic used, threshold for that statistics and both the coarse and fine banks used. Then for each data segment and the corresponding fine subbank, we perform a search with full sampling rate of $4096$ Hz as used in single stage flat search. Then we follow the same procedure as the first stage by again collecting single detector triggers from the fine subbank and obtaining the second stage coincident triggers by matching parameters etc. We then cluster these coarse and fine subbank triggers together and obtain the final triggers. Now these final triggers need to be compared with the noise background for estimating their statistical significance. For this second stage, we use the $\rho_{\rm single, II}$ which is the same as that for the flat search. To estimate the noise background, we have used the single detector trigger time slides but with some caveats which we will discuss in the next section.
We now present our results.

\section{Comparison with the flat search}
\label{hierarchical_result}

In this section, we compare the results for the hierarchical search with the  flat search. For this analysis we assume stationary Gaussian noise. We begin by comparing the noise background and noise foreground without injections. For each individual detector, on an average, we found $\sim ~53$ triggers per second from the flat search with $\rho_{\rm single, flat} = 5.5$ and $\sim ~7 - 8$ triggers per second from the coarse search with $\rho_{\rm single, I} ~=~ 5.5$. But for $\rho_{\rm single, I} ~=~ 5.0$, we obtained $\sim ~111$ triggers per second. We use full banks for both flat and stage I to obtain the triggers. We observe reduced number of triggers with the coarse bank although we have kept the same threshold. This is because of the reduced number of the templates in the coarse bank and also the reduced sampling rate. We expect that with only quarter of the templates and $1/8$th of the sample points, we can at best get a factor of $\sim ~32$ reduction in the computational cost as  that for the flat search. This is because the main cost of the search comes from  matched filter computations and is due to FFT operation used in the  data analysis. The cost of a FFT scales as $N\log{N}$ where $N$ is the number of data points in the data segment. We have about quarter the number of  matched filter computations because of the reduced number of the templates in the coarse bank and also the cost of each FFT goes down by the factor of about 8 due to fewer data points per segment due to reduced sampling rate. But actual speed-up depends on the coarse stage threshold $\rho_{\rm single, I}$ which will in turn, decide the number of stage ${\rm I}$ candidate triggers per data segment that need to be followed up. Also, the computational cost depends on size of the stage ${\rm II}$ fine subbank which depends upon the choice of the relevant neighbourhood for each stage ${\rm I}$ trigger template of the coarse bank. For instance, if we apply a  very low single detector threshold e.g. $\rho_{\rm single, I} \sim ~3.5$, we will get huge number of coincident stage ${\rm I}$ triggers from the entire  coarse bank and also from the full parameter space. This will entail employing almost all of the fine bank because we will need the union of  individual coarse trigger template neighbourhoods (the fine subbank will become almost the fine bank) and we may lose on the computational benefits arising from the hierarchical strategy.

\begin{figure}[h]
\begin{center}
\includegraphics[width=0.5\textwidth]{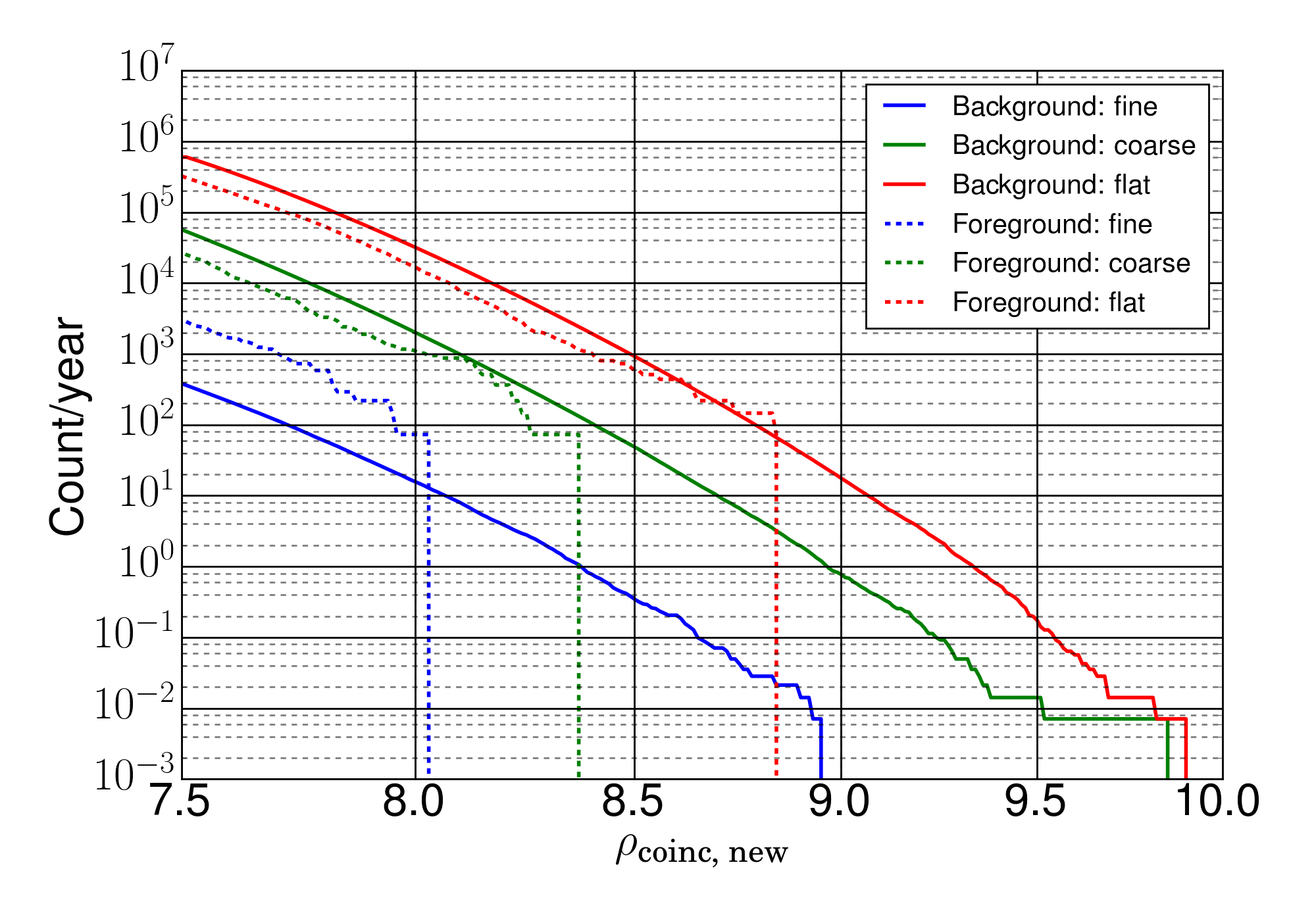}
 \caption{Full search background and foreground event rates (per year). The foreground data is of $\sim$ 5 days and background amounts to more than 140 years after time slides.}
\label{background_per_year}
\end{center}
\end{figure}

\begin{figure}[h]
\begin{center}
\includegraphics[width=0.5\textwidth]{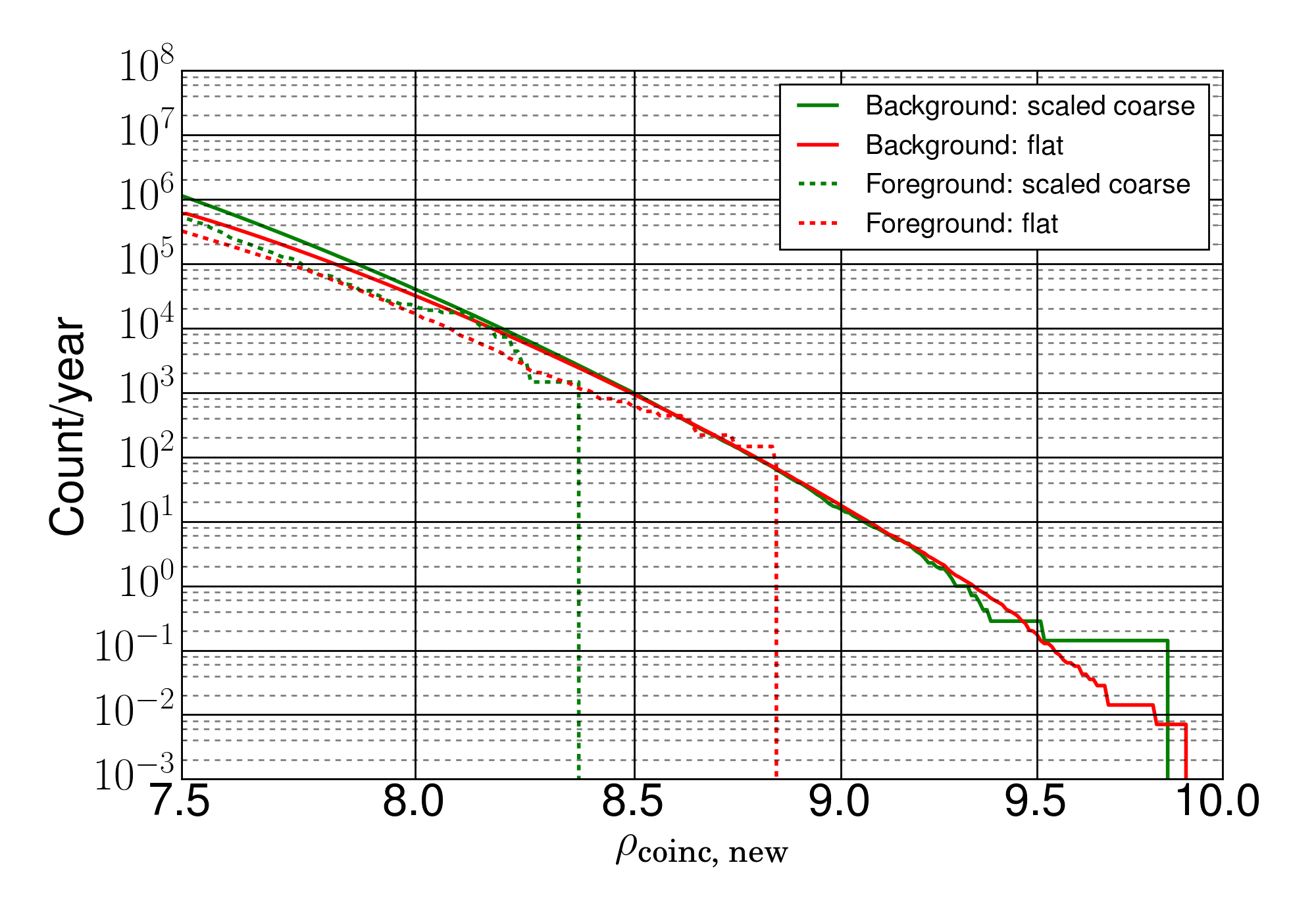}
\caption{Full search background and foreground event rates (per year). The first coarse stage rates are scaled by the factor of $\sim$ 20 to match all the events. The scale factor to get back the same event rates is same as the speed-up factor for the simulated data.}
\label{scaled_background_per_year}
\includegraphics[width=0.5\textwidth]{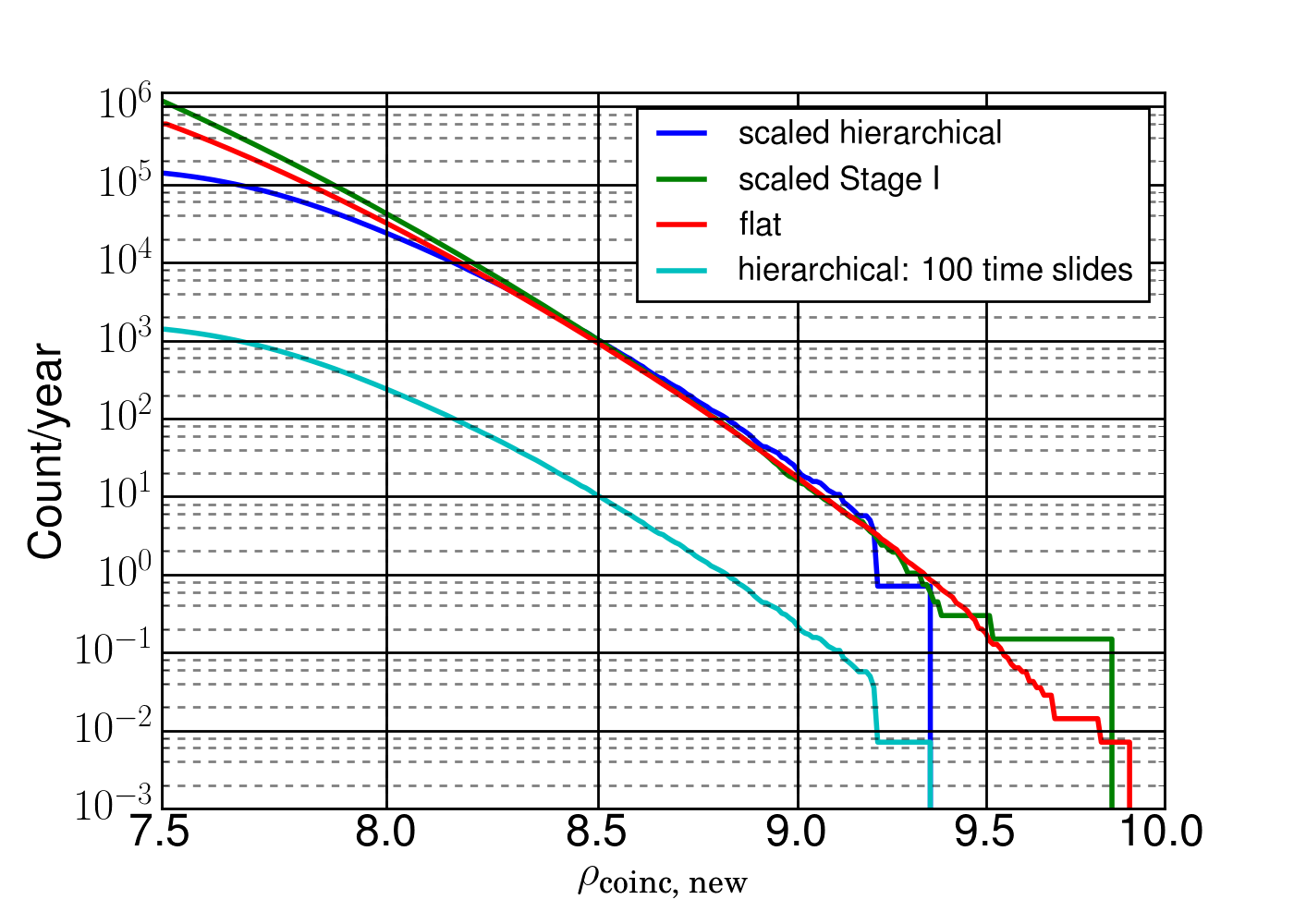}

\caption{Figure shows the flat search background (red), Stage I background scaled by the speed-up factor (green) and hierarchical search (flat search equivalent) background calculated using 100 time-slides (blue) and scaled with the constant factor to get $\sim$ 141 years duration background. It can be see that both the scale backgrounds match well with the flat search background. 100 time-slides require the Stage II to matched filter the data segment with $> 30000$ fine bank templates.}

\label{scaled_hbackground_per_year}
\end{center}
\end{figure}

\begin{figure}[h]
\begin{center}
\includegraphics[width=0.5\textwidth]{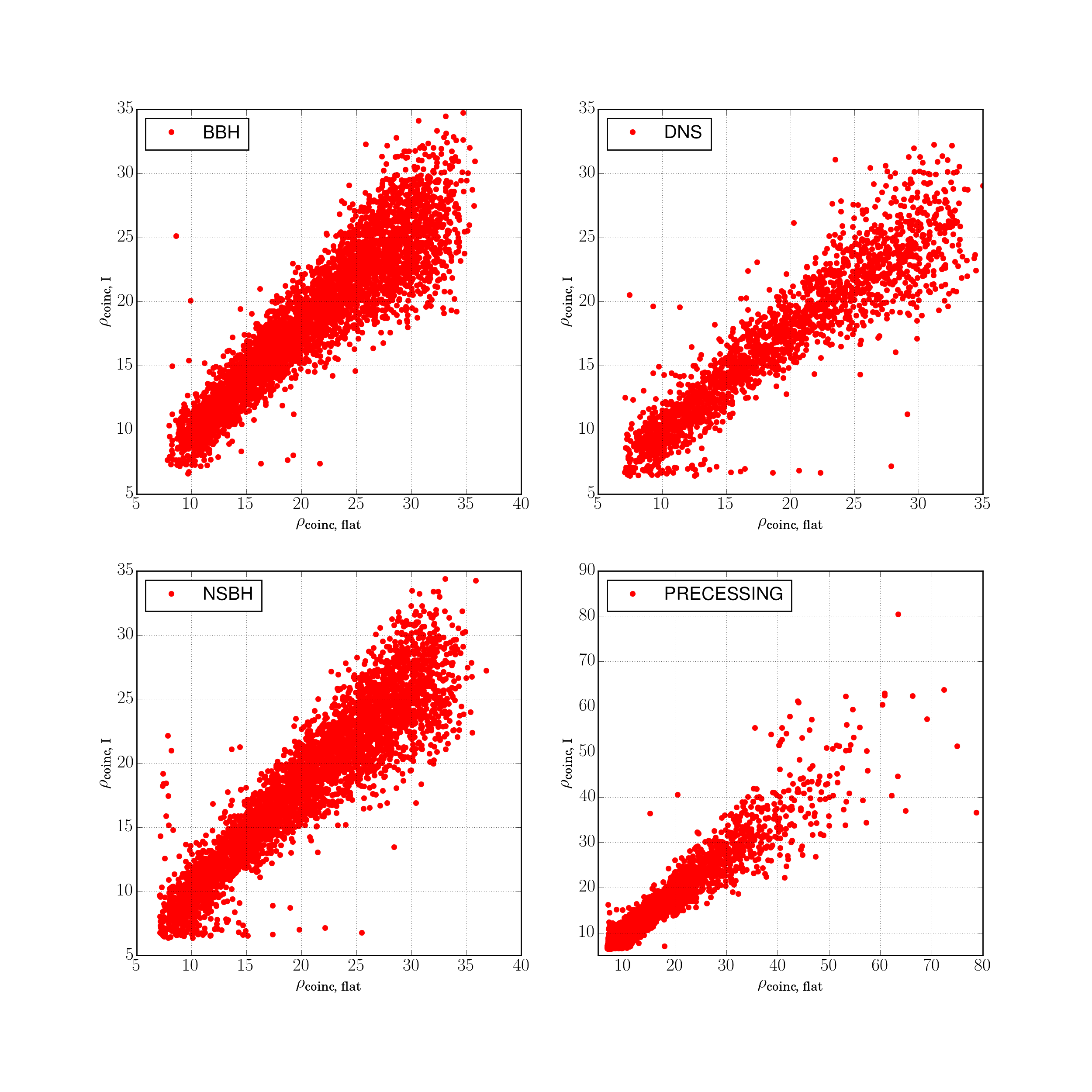}
 \caption{Coincident recovered newSNRs for all the injections.  It shows how much SNR is lost in stage I with the lower sampling rate and the coarse bank which stage II may recover with fine subbank and full sampling rate.}
\label{all_inj_coinc_coarse_vs_flat}
\end{center}
\end{figure}

\begin{figure}[h]
\begin{center}
\includegraphics[width=0.5\textwidth]{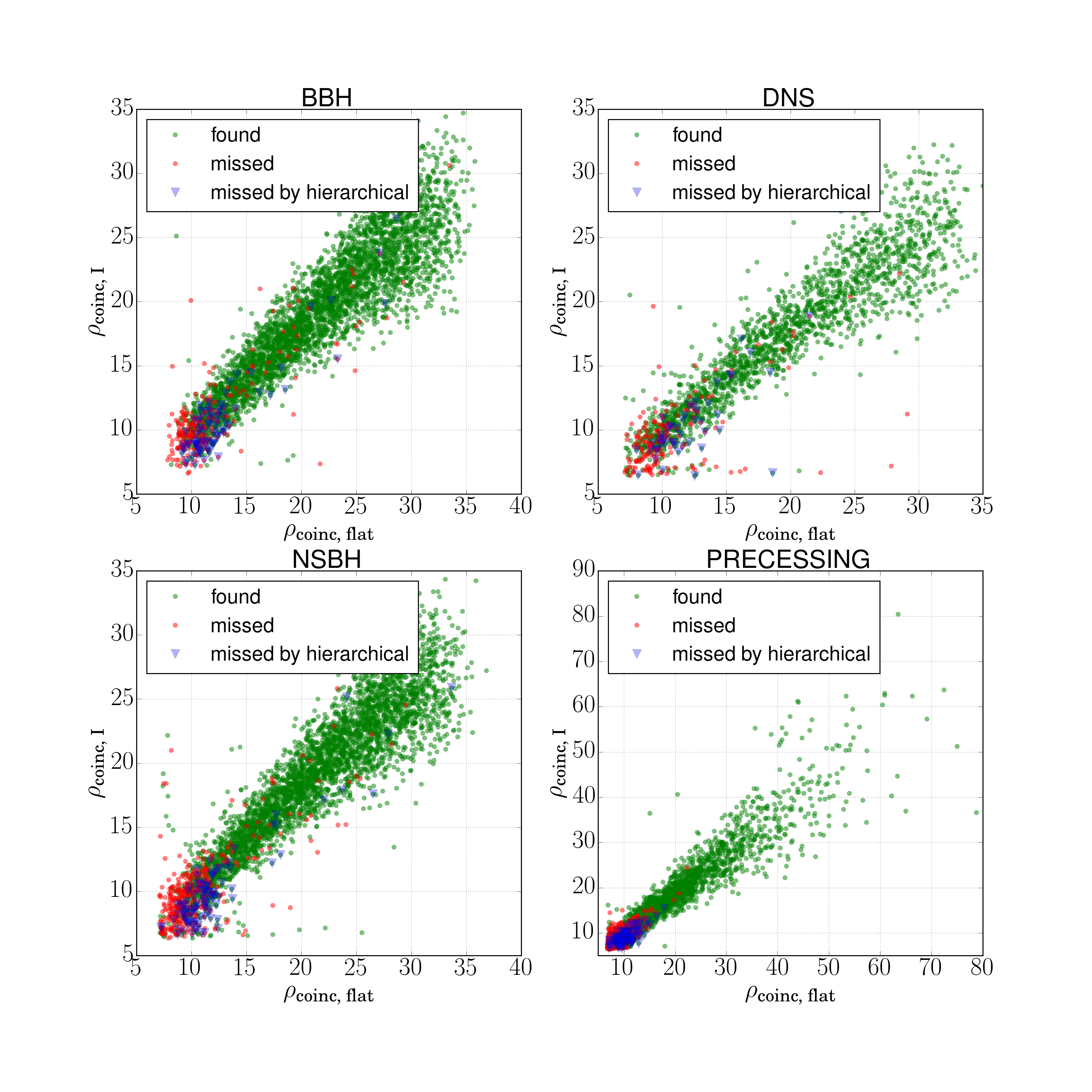}
 \caption{The plot shows recovered coincident newSNR for the signals missed and found simultaneously by the hierarchical search with red and green dots respectively. The blue triangles shows the injections missed by the hierarchical search but found by the flat search. These are $\sim 2\%$ of the total injections.}
\label{all_inj_coinc_coarse_vs_flat}
\end{center}
\end{figure}

With the help of time slides, we compute the noise background for more than $140$ years of coincident data. This is done for the flat search in the usual way and also for both the stages of the hierarchical search which uses the coarse bank and the zero-lag fine subbanks. Foreground is computed for 5 days of the coincident data for the flat and the 2-stage hierarchical search. Note that the thresholds and other parameters (clustering etc.) used for the background, foreground and injection recovery in stage ${\rm II}$ of the hierarchical search are the same as that of the flat search. Figure~\ref{background_per_year} shows backgrounds and foregrounds for the flat search, stage ${\rm I}$ and stage ${\rm II}$ in terms of cumulative  number of coincident events per year with newSNR plotted on the horizontal axis. The figure shows that stage ${\rm I}$ background and foreground is lower by almost an order of magnitude than that of the flat search. The hierarchical search background is the union of the coarse background obtained in stage ${\rm I}$ and the fine subbank background obtained from stage ${\rm II}$. The figure~\ref{background_per_year} shows that the contribution to the total hierarchical background from stage ${\rm II}$ is almost negligible. So hierarchical background is practically same as that of stage ${\rm I}$ background. However, stage ${\rm II}$ foreground contributes comparatively much more to the overall foreground. This is because we use fine subbanks in stage ${\rm II}$ constructed using only `0'-lag (foreground) coincident triggers from stage ${\rm I}$. All templates in the subbank can contribute to the foreground evaluation which give better chances for noise coincidences. But, when the background for stage ${\rm II}$ is calculated using time slides, the single detector triggers from from two different data segments are likely to have very few or no common templates in the stage ${\rm II}$ fine subbank. This may lead to some bias in the total hierarchical background but it should be negligible as the figure \ref{background_per_year} shows. So, in practise, we can use the stage ${\rm I}$ background to assign statistical significance to the triggers and it can be simply scale to obtain the flat search background for the simulated data. We can, in principle, estimate the background which is equivalent to the flat search background using time slides. But then we lose all the computational benefits of the hierarchical search. Even with $< 2000$ time slides, the data segment involved, go through the match over the full fine bank in stage II of the hierarchy. This utilisation of the full fine bank in stage II implies that when we do time slides with the non-zero lag triggers from stage I, we recover the same background as the flat search. Hence, at least with the simulated coloured detector noise, we came up with the idea of scaling the stage I background calculated using the coarse bank to recover the flat search background. Interestingly, if we scale the stage I background by the speed-up factor, we recover the flat search background over the coincident new-SNR threshold of 8. This can be seen in figure~\ref{scaled_background_per_year}. Moreover, same scaling factor does match the noise-only foreground of stage I of the hierarchy with that of the flat search. A little excess in the low new-SNR region between 7.5 to 8 is the reminiscent effect of the reduced single detector and coincident thresholds for the stage I.  The same scaling as the speed-up factor works for the stationary Gaussian data is due to the fact that the same scale determines the number of random variables used as matched filter statistic to get the noise-only back and foregrounds. This speed-up factor is explained in detail later. {\em Thus, in the simulated data case, we can use parameter space independent scaling argument to scale correctly stage I background to get the equivalent flat search background. Using this equivalent background, we an assign a similar significance to the foreground triggers.} Also we estimate flat search equivalent background by using non-zero lag 100 time-slides corresponding to 1.4 years of the background. This is shown in Figure~\ref{scaled_hbackground_per_year} by the magenta line. If we scale the non-zero lag hierarchical background to get the background equivalent of $\sim 141$ years, we can see that the scaled hierarchical search background matches well with the flat search and the scaled stage I background except for the low SNR region. Even just 100 non-zero lag time slides need to compute matched filters with more than 30000 fine bank templates in stage II for the each of the data segment. With real data, this exercise needs to be carried out even more cautiously to see if we can recover the flat background and if the scale factor is more or less the speed-up factor for the real data. Then we can use this method to get the flat equivalent background without losing any of the computational advantages we get by using the hierarchical strategy as the speed-up factor can easily be evaluated by running flat searches on small parts of the data. This question needs more detailed and thorough investigation with real data.

Given the above description, we get the background for $\rho_{\textrm{I, coinc}} \sim 9.3$ and $\rho_{\textrm{flat, coinc}} \sim 9.7$ for a false alarm rate of 1 per 50 yr for the hierarchical and flat search respectively (See Fig. (\ref{background_per_year})). This opens up the possibility of detecting few more borderline GW signals having $\rho_{\textrm{coinc}} ~\in~ (9.3, 9.7)$. The number of injections recovered with the newSNR more than above mentioned thresholds are given in the table~\ref{table:found_inj}. If we can recover the full SNR using hierarchical search without losing the signals that are found by the flat search, we may detect few more signals dew to the reduced background. For the current study, we have not put any threshold on the coincident statistic for obtaining stage ${\rm I}$ coincident triggers. We follow up all the coincident triggers from the first stage. We may opt to apply the coincident threshold for real data if we obtain too many first stage coarse triggers per segment or if we use lower individual detector threshold $\rho_{single, I}$. Varying this threshold  decides the speed up and efficiency of the hierarchical search over the flat search.

\begin{table}
\label{table:found_inj}
\begin{center}
 \begin{tabular}{||c | c | c | c||}
 \hline
 Injections & Total & Coarse $> 9.3 $ & Flat $> 9.7 $ \\ 
 \hline\hline
 BBH & 4342 & 3907 & 3881 \\
 \hline
 NSBH & 4342 & 3912 & 3890 \\
 \hline
 DNS & 2171 & 1860 & 1835 \\
 \hline
 Precessing & 8684 & 2520 & 2423 \\ 
 \hline

 \hline

\end{tabular}
\end{center}
\caption{Number of injections as detected by Stage I of the hierarchical search and the flat search with the same FA of 1 per 50 yrs as per the search.}
\end{table}

\begin{figure}[h]
\begin{center}
\includegraphics[width=0.475\textwidth]{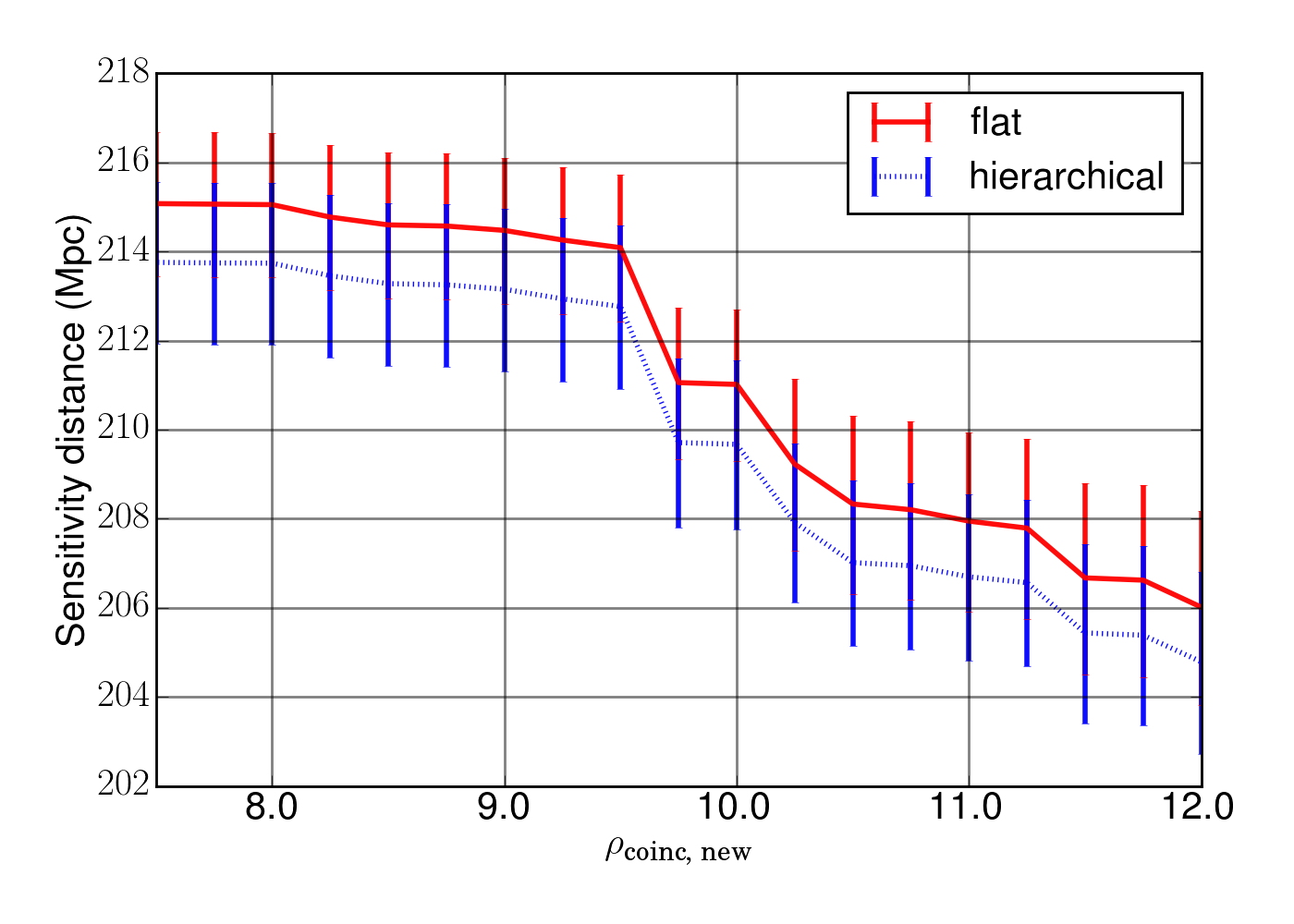}
\includegraphics[width=0.475\textwidth]{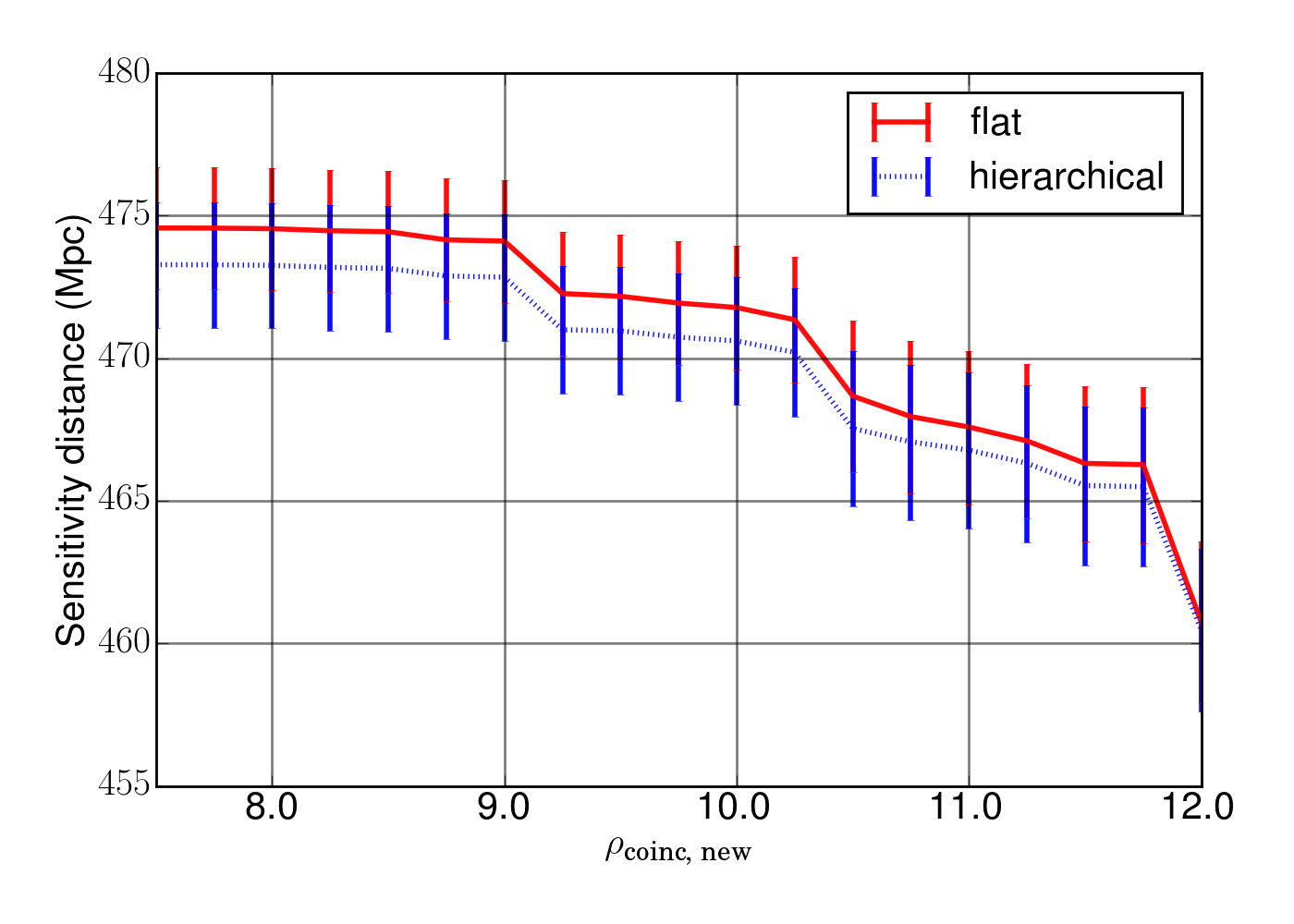}
\caption{Comparison of searches for distance sensitivity. Left:aligned-spin DNS injections and right: aligned-spin NSBH injections}
\label{sensitivity_distance1}
\end{center}
\end{figure}

\begin{figure}[h]
\begin{center}
\includegraphics[width=0.475\textwidth]{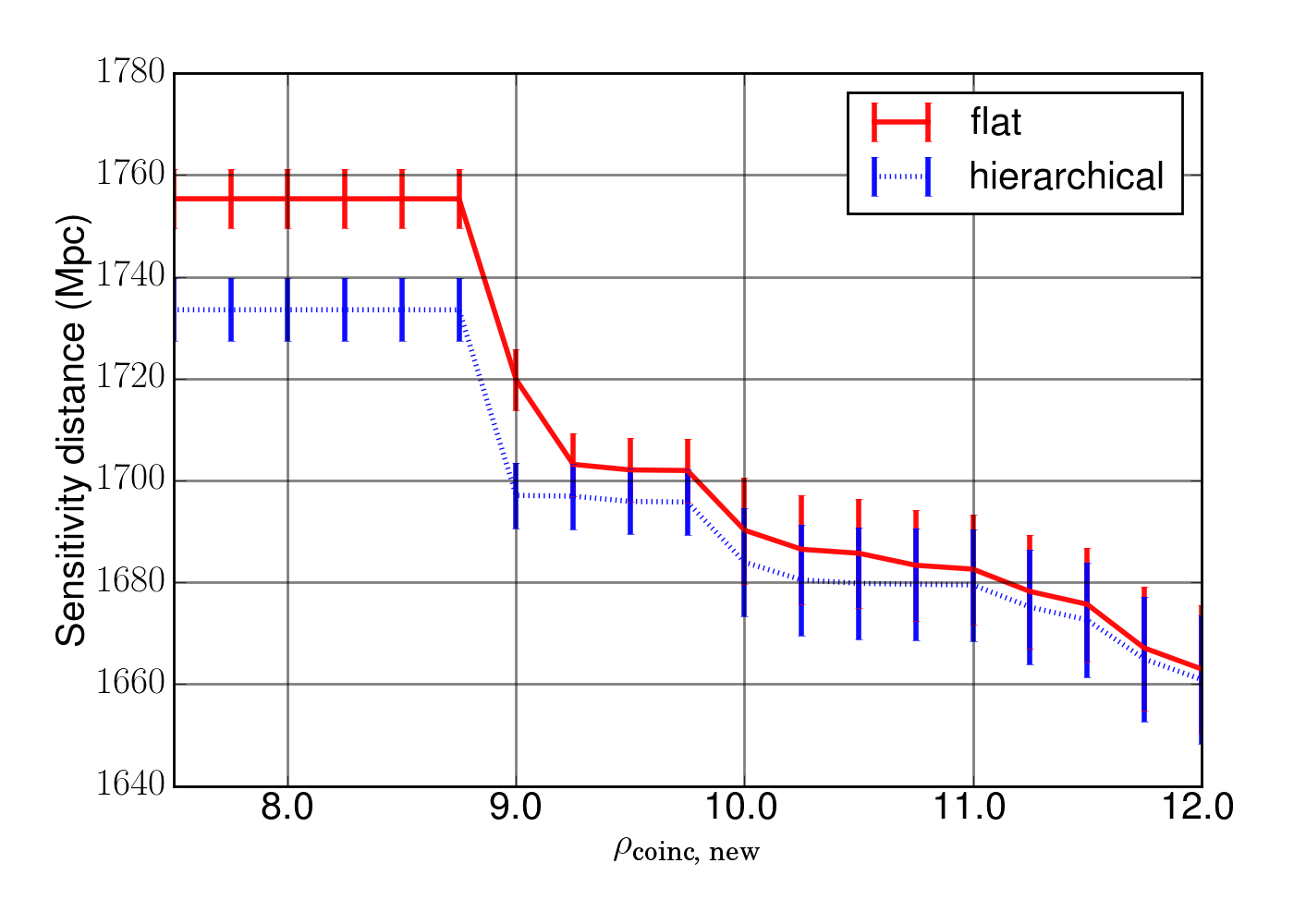}
\includegraphics[width=0.475\textwidth]{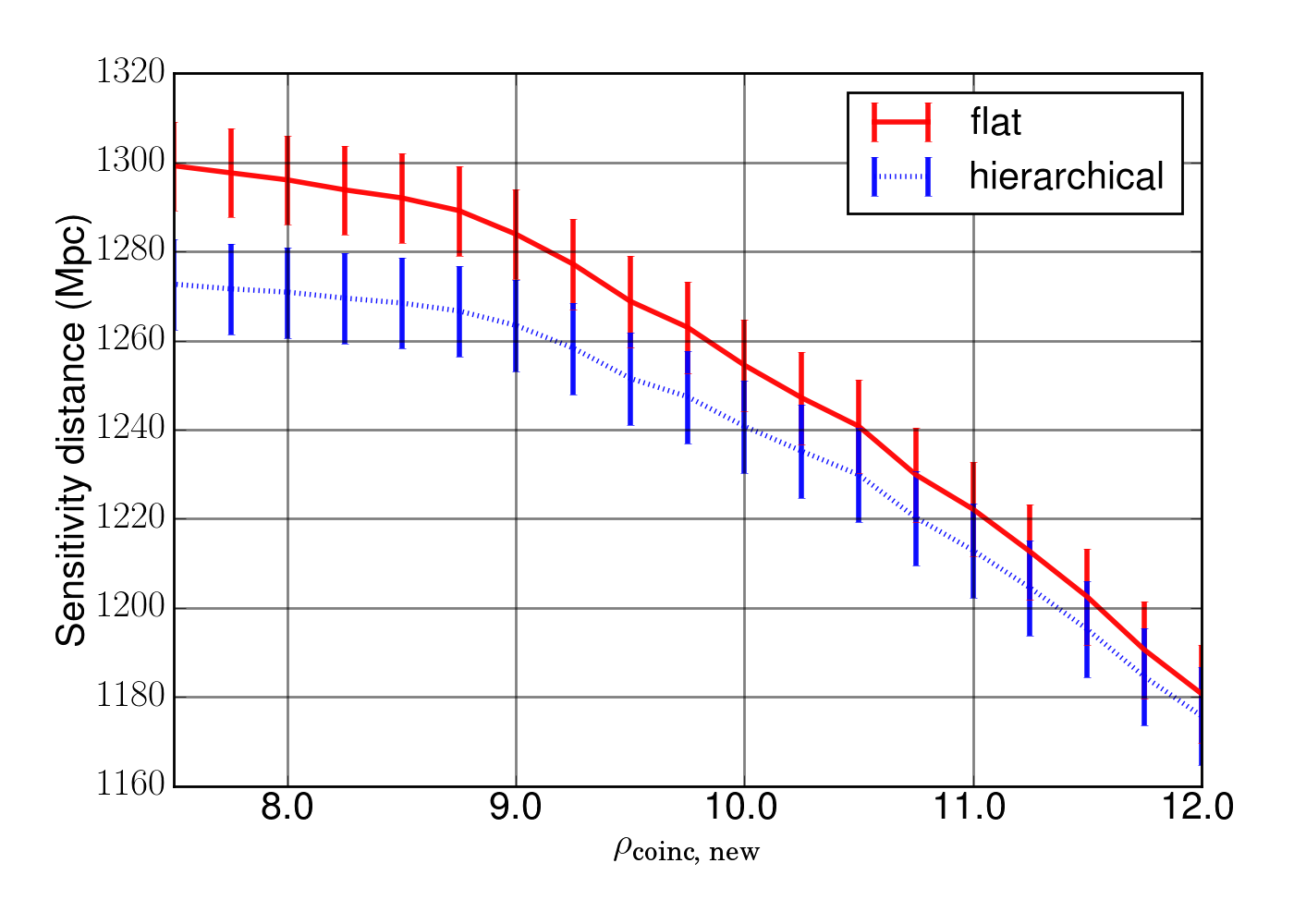}
\caption{Comparison of searches with distance sensitivity: top:aligned-spin BBH injections and bottom: precessing injections}
\label{sensitivity_distance2}
\end{center}
\end{figure}

\begin{figure}[h]
\begin{center}
\includegraphics[width=0.5\textwidth]{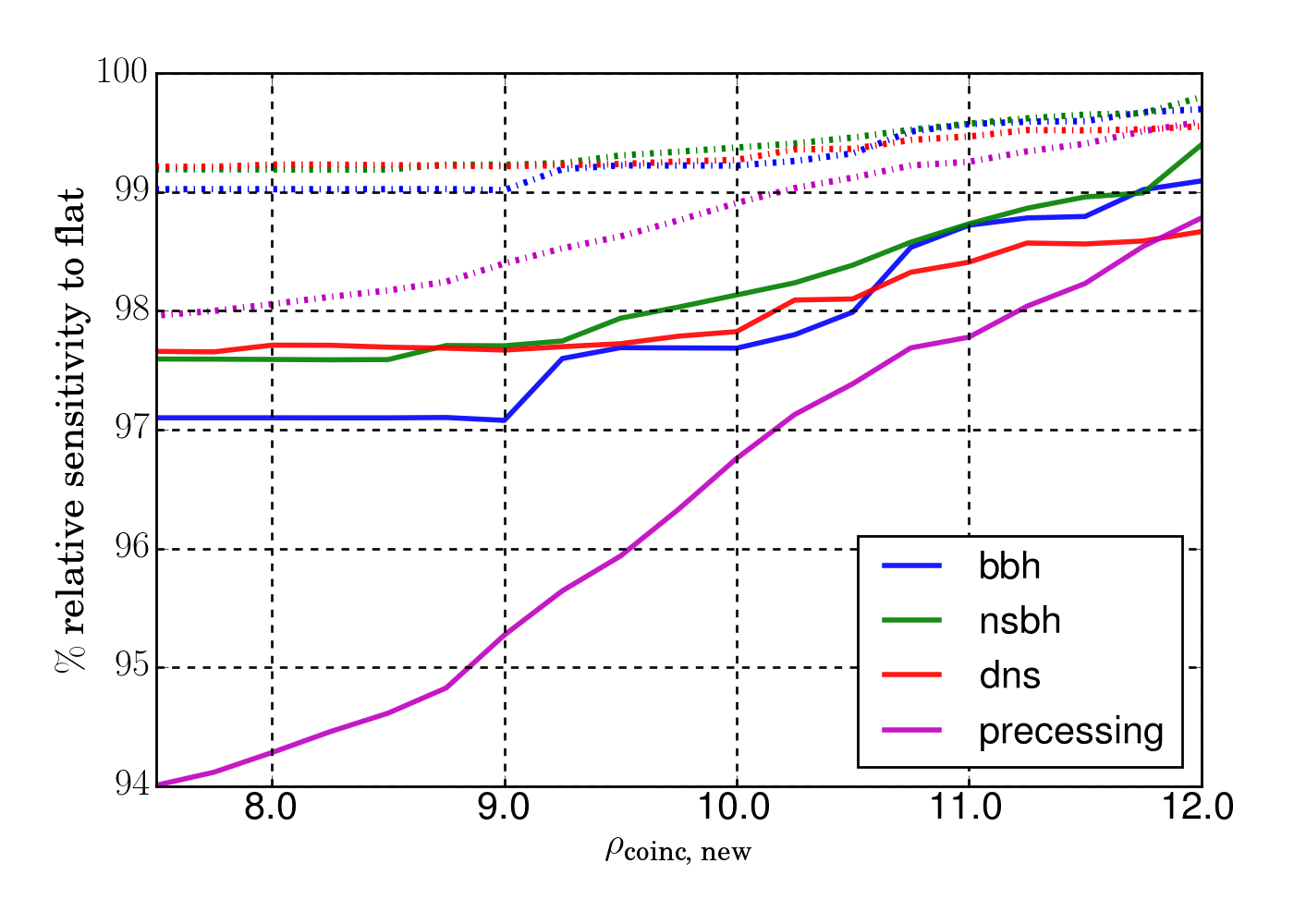}
\caption{Plot shows the relative sensitivity (in \% ) of the hierarchical search with respect to the flat search. Solid lines show relative sensitive volume and dash-dotted lines show relative sensitivity distance.}
\label{sensitivity_distance_compare}
\end{center}
\end{figure}
We now investigate the recovery of injected CBC signals by both types of searches, flat and hierarchical. As discussed in section ~\ref{hierar_aligned_spin}, we inject 10000 aligned spin CBC (DNS, NSBH and BBH) signals. In addition, we inject more than 8000 precessing CBC (NSBH and BBH) ones. Figure~\ref{all_inj_coinc_coarse_vs_flat} shows all the above mentioned injections as missed and found. For each of the subplot, we have plotted injection with flat coincident SNR against the coincident newSNR for the stage I of the hierarchy. The red dots show the injections missed by the both flat and hierarchical search while green dots show the injections found by both the searches. The blue triangles denote the injections missed only by the hierarchical search but found by the flat search. For all the 3 aligned spin cases, the injections missed only by the hierarchical search are $\sim 2\%$ of those found by the flat search. For the precessing case, hierarchical search loses $\sim 6\%$ of the injections recovered by the flat search. Figures~\ref{sensitivity_distance1} and ~\ref{sensitivity_distance2} show sensitivity distance for both searches, hierarchical and the flat, with a varying coincident newSNR threshold. The newSNR threshold corresponds to a false alarm rate as can be seen from the figure~\ref{background_per_year}. We see that both, hierarchical and flat, searches have almost similar sensitivity distances as a function newSNR or as a function of the false alarm rate as shown in Figure~\ref{sensitivity_distance_compare}. This implies that both the searches perform almost equally well. For the calculation of the sensitivity distance, we have used all the CBC injections. It can be seen that only the DNS search has slightly lower sensitivity for the hierarchical search than that for the  flat search. This is expected as we are using truncated waveforms with much lower MM and BNS signals are of long duration and contribute significantly to the SNR at higher frequencies which means that the fractional loss in SNR is more. The lower recovery of signals is due to reduced stage ${\rm II}$ SNR compared to flat search for a few BNS sources. This is because the SNR of these sources in the one of the detectors is slightly lower so that some false trigger templates are contributed from the stage ${\rm I}$. Thus we see that, the hierarchical search recovers almost all the injections as those  recovered by the flat search. There is a slight advantage to the hierarchical search over the flat search, because we can choose a slightly lower detection threshold with hierarchical search for the same false alarm rate. We have not addressed this question here. Thus for the hierarchical procedure we have proposed, we conclude that both the searches have almost similar distance  sensitivity for the injected set of signals. Next we consider the computational cost of each kind of search.

We now look more closely at the computational costs and the computational gain from the hierarchical search. We also explain how the same is related to the background estimation. For a data segment of length 256 sec, we have few hundreds of coincident first stage triggers (no additional coincident threshold is applied). On an average, we have 40-90 templates in the neighbourhood of the each of the coarse trigger template as can be seen in the figure~\ref{fine_bank_nhbd}. We then obtain a second stage fine subbank which is the union of neighbourhoods. This subbank has 1000 - 4000 templates on an average per data segment. We now compute the average number of Floating Point Operations (FLO) per data segment. We do 60000 MF calculations at 512 Hz sampling rate in the first stage and at most 4000 MF calculations at 4096 Hz sampling rate in the second stage. Each MF computation involves a complex FFT corresponding to the two phases of the waveform. On the other hand, in the flat search, we do 250,000 MF calculations at the sampling rate of 4096 Hz. Each MF calculation uses Discrete Fourier Transform (DFT). Each DFT with N data points requires $\alpha N \log{N}$ FLO, where $\alpha \sim 3$ for a real DFT and double this number for a complex FFT and depends on the algorithm used. Thus, roughly, discarding the $\alpha$ factor which is common to both the searches, the flat search requires $250 \times 4.096$ mega-FLO while the hierarchical search strategy adopted here, requires $60 \times 0.512 ~+~ 4 \times 4.096$ mega-FLO. Thus one obtains a computational gain of $\sim$ 20.

Now we look back at the estimation of the noise background for the hierarchical search. We argue that the hierarchical background is just scaled down from the flat search background roughly by the speed up factor, which in this case is $\sim 20$. The noise background arises from the number of triggers which essentially stem from the number of independent Gaussian random variables in the matched filter output. The Gaussian variables in the matched filter output are however correlated. For the flat search we get roughly $256 \times 250000 \times 4096$ data points (Gaussian variables not necessarily independent) per segment. But for the hierarchical search we must consider both Stage I and Stage II data points. For the hierarchical search we have $256 \times 60000 \times 512$ + $256 \times 4000 \times 4096$ data points per segment. We may expect the effect of correlation between Gaussian variables to be about the same in both flat case and the hierarchical case. Ignoring the effect correlations and except for the slowly varying factor of $\log N$, the ratio of independent Gaussian variables in the two situations is roughly the same as the ratio of matched filtering operations required for each of the searches. This is in fact the speed up factor. This is evident from figures~\ref{background_per_year} and ~\ref{scaled_background_per_year}. However real data contains non-Gaussian artefacts and we basically sample the tail of the noise distribution (rare events) to estimate the background. Therefore, this scaling exercise needs to be carried out carefully in order to obtain the correct scaling. The scaling may depend upon template duration as very short duration templates are more susceptible to the glitches and artefacts in the real data.

We now make a few remarks. First of all, the non-precessing injections we used are in the H1-L1 coincident SNR range $8$ to $30$ and our precessing injections are linearly distributed in distance. Secondly, we get our noise background for the hierarchical search almost only from the first stage of the hierarchy. In order to obtain the full background equivalent to the flat search time slide background, the second stage fine subbank must be obtained with non-zero lag stage {\rm I} coincident triggers. With sufficiently large  number of the time slides the union of fine subbank over time slides will be almost as large as the whole fine bank employed in the flat search. This will compromise all the computational advantage that one expects to get by the hierarchical search.

We may, therefore, be able to use the stage ${\rm I}$ background to infer the significance of detection, after further investigations with real data. This may not be exactly equivalent to the flat search background, but it can be used as a separate hierarchical background. Otherwise, we can do few hundred time-slides to get the noise background for shorter duration of the data and then scale it to estimate the full flat search background to get the full significance. We propose to address this background issue and also tune the pipeline for injection recovery per mass bin with the real data in the future work.

\section{Discussions and Future prospects}
\label{discussion}
In this work, we have demonstrated that the two stage hierarchical search on simulated data containing Gaussian stationary noise is $\sim 20$ times faster than the current flat search used for LIGO O1 analysis. This factor of reduction in computational cost has been obtained without any optimisation. With a judicial choice of parameters we have shown that it can be almost as good as the single stage flat search in sensitivity - that is given a set of injections, this search detects as many signals as the flat search. In future, we propose to run and optimise our 2-stage hierarchical search on O1 data and examine how it performs. We expect that non-Gaussian detector data containing glitches can produce a large number of false alarms which may reduce the performance of the hirerchical search more compared to the flat search.

As pointed out before, the computational effort saved by doing a hierarchical search can be used elsewhere. It can be used to do more detailed analysis of the detected CBCs such as test of general theory of relativity by comparing waveforms predicted by other theories of gravity etc. The saved CPU time could be used to search for other astrophysical sources. This issue will become all the more important when detectors become more sensitive in the future. The demand for computation will increase because  the event rate will go up with the corresponding requirement of a much denser template bank covering the parameter space.

The 2-stage hierarchical method can be readily employed for online searches where we do not worry about assigning the exact significance using full background estimation. We can use the false alarm rate generated by the first stage to get online triggers much faster owing to speed up we get due the hierarchical algorithm.

Another important direction to follow is the implementation of a hierarchical search with precessing waveforms. We believe that the order of magnitude reduction in the computational cost will allow us to do at least partial precessing searches. But creating a template bank with precessing templates is also very difficult as it has to be done stochastically~\cite{Harry:2016ijz}. We plan to explore the possibility of performing multi-stage hierarchical searches using hybrid (non-precessing + partial precessing) template banks.

\begin{acknowledgments}
We would like to thank Albert Lazzarini (Caltech)  for suggesting us the original idea. We like to acknowledge Alex Nitz (AEI-Hannover), Remya Nair (Kyoto University) for reading the manuscript and suggestions. We would also like to thank Anand Senpupta (IITG), Sukanta Bose (IUCAA), Badri Krishnan (AEI-Hannover), Ian Harry (AEI-Potsdam), Sumit Kumar (ICTS) and Patrick Brady (UWM) for the useful discussions. We acknowledge the use of IUCAA LDG cluster Sarathi for the computational/numerical work. BUVG acknowledges the support of University Grants Commission (UGC), India. This research benefited from a grant awarded to IUCAA by the  Navajbai Ratan Tata Trust (NRTT). S. M. acknowledges support from the Department of Science \& Technology (DST), India provided under the Swarna Jayanti Fellowships scheme. SVD  acknowledges the support of the Senior Scientist NASI Platinum Jubilee Fellowship.
\end{acknowledgments}

\bibliography{hierarchical}

\end{document}